\documentclass{elsarticle}

\usepackage[reviewcopy]{adndt}
\usepackage{longtable}

\usepackage{amsmath}

\usepackage{amssymb}

\biboptions{square,sort&compress}
\bibpunct[]{[}{]}{,}{n}{}{;}
\usepackage{color}

\begin{document}

\begin{frontmatter}

\journal{Atomic Data and Nuclear Data Tables}


\title{Discovery of Chromium, Manganese, Nickel, and Copper Isotopes}


\author{K. Garofali}
\author{R. Robinson}
\author{M. Thoennessen\corref{cor1}}\ead{thoennessen@nscl.msu.edu}

\cortext[cor1]{Corresponding author.}

\address{National Superconducting Cyclotron Laboratory and \\ Department of Physics and Astronomy, Michigan State University, \\ East Lansing, MI 48824, USA}

\begin{abstract}
Twenty-seven chromium, twenty-five manganese, thirty-one nickel and twenty-six copper isotopes have so far been observed and the discovery of these isotopes is discussed. For each isotope a brief synopsis of the first refereed publication, including the production and identification method, is presented.
\end{abstract}

\end{frontmatter}





\newpage
\tableofcontents
\listofDtables

\vskip5pc

\section{Introduction}\label{s:intro}

The discovery of the chromium, manganese, nickel and copper isotopes is discussed as part of the series summarizing the discovery of isotopes, beginning with the cerium isotopes in 2009 \cite{2009Gin01}. Guidelines for assigning credit for discovery are (1) clear identification, either through decay-curves and relationships to other known isotopes, particle or $\gamma$-ray spectra, or unique mass and Z-identification, and (2) publication of the discovery in a refereed journal. The authors and year of the first publication, the laboratory where the isotopes were produced as well as the production and identification methods are discussed. When appropriate, references to conference proceedings, internal reports, and theses are included. When a discovery includes a half-life measurement the measured value is compared to the currently adopted value taken from the NUBASE evaluation \cite{2003Aud01} which is based on the ENSDF database \cite{2008ENS01}. In cases where the reported half-life differed significantly from the adopted half-life (up to approximately a factor of two), we searched the subsequent literature for indications that the measurement was erroneous. If that was not the case we credited the authors with the discovery in spite of the inaccurate half-life.

The first criterion is not clear cut and in many instances debatable. Within the scope of the present project it is not possible to scrutinize each paper for the accuracy of the experimental data as is done for the discovery of elements \cite{1991IUP01}. In some cases an initial tentative assignment is not specifically confirmed in later papers and the first assignment is tacitly accepted by the community. The readers are encouraged to contact the authors if they disagree with an assignment because they are aware of an earlier paper or if they found evidence that the data of the chosen paper were incorrect.

The second criterion affects especially the isotopes studied within the Manhattan Project. Although an overview of the results was published in 1946 \cite{1946TPP01}, most of the papers were only published in the Plutonium Project Records of the Manhattan Project Technical Series, Vol. 9A, ''Radiochemistry and the Fission Products,'' in three books by Wiley in 1951 \cite{1951Cor01}. We considered this first unclassified publication to be equivalent to a refereed paper.

The initial literature search was performed using the databases ENSDF \cite{2008ENS01} and NSR \cite{2008NSR01} of the National Nuclear Data Center at Brookhaven National Laboratory. These databases are complete and reliable back to the early 1960's. For earlier references, several editions of the Table of Isotopes were used \cite{1940Liv01,1944Sea01,1948Sea01,1953Hol02,1958Str01,1967Led01}. A good reference for the discovery of the stable isotopes was the second edition of Aston's book ``Mass Spectra and Isotopes'' \cite{1942Ast01}.

\section{Discovery of $^{42-68}$Cr}

Twenty-seven chromium isotopes from A = $42-68$ have been discovered so far; these include 4 stable, 9 proton-rich and 14 neutron-rich isotopes.  According to the HFB-14 model \cite{2007Gor01}, $^{77}$Cr should be the last odd-even particle stable neutron-rich nucleus while the even-even particle stable neutron-rich nuclei should continue through $^{84}$Cr. At the proton dripline $^{40}$Cr and $^{41}$Cr should still be particle stable. Thus, about 15 isotopes have yet to be discovered corresponding to about 36\% of all possible chromium isotopes.

\begin{figure}
	\centering
	\includegraphics[scale=.5]{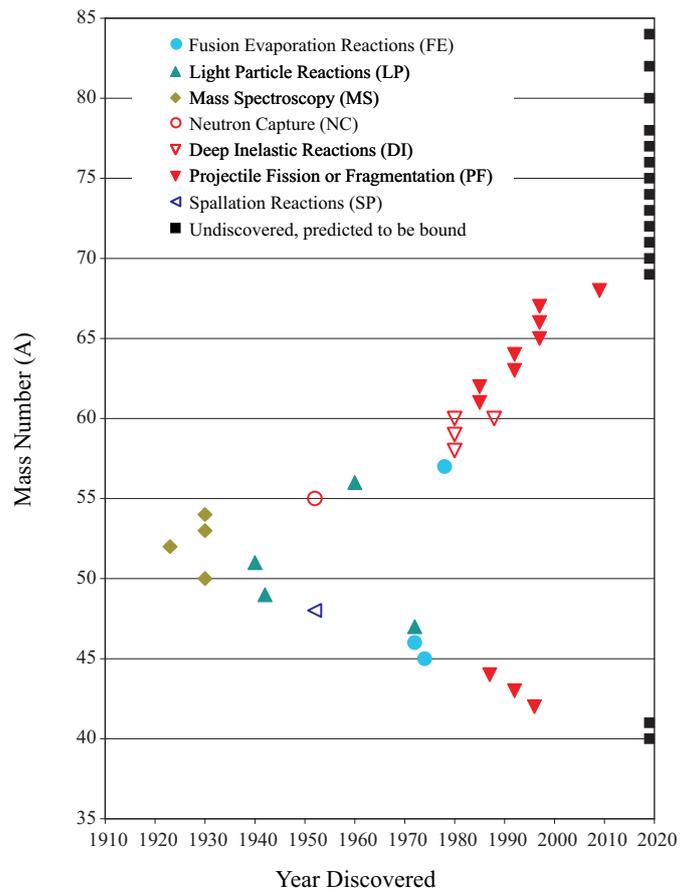}
	\caption{Chromium isotopes as a function of time when they were discovered. The different production methods are indicated. The solid black squares on the right hand side of the plot are isotopes predicted to be bound by the HFB-14 model.}
\label{f:year-cr}
\end{figure}

Figure \ref{f:year-cr} summarizes the year of first discovery for all chromium isotopes identified by the method of discovery. The range of isotopes predicted to exist is indicated on the right side of the figure. The radioactive chromium isotopes were produced using deep-inelastic reactions (DI), light-particle reactions (LP), neutron capture (NC), spallation (SP), heavy-ion fusion evaporation (FE) and projectile fragmentation of fission (PF). The stable isotopes were identified using mass spectroscopy (MS). Heavy ions are all nuclei with an atomic mass larger than A=4 \cite{1977Gru01}. Light particles also include neutrons produced by accelerators. In the following, the discovery of each chromium isotope is discussed in detail and a summary is presented in Table 1.

\subsection{$^{42}$Cr}\vspace{0.0cm}
In the paper ``First Observation of the T$_{z}=-$7/2 Nuclei $^{45}$Fe and $^{49}$Ni'', Blank et al. reported the discovery of $^{42}$Cr in 1996 at the Gesellschaft f\"{u}r Schwerionenforschung (GSI) in Germany \cite{1996Bla01}. A 600 A$\cdot$MeV $^{58}$Ni beam bombarded a beryllium target and isotopes were separated with the projectile-fragment separator FRS. $^{42}$Cr was identified by time-of-flight, $\Delta$E, and B$\rho$ analysis. ``We observed ten events of $^{42}$Cr, three events of $^{45}$Fe, and five events of $^{49}$Ni. These three isotopes have been identified for the first time in the present experiment.''

\subsection{$^{43}$Cr}\vspace{0.0cm}

$^{43}$Cr was discovered by Borrel et al. at the Grand Accelerateur National D'ions Lourds (GANIL) in France in 1992, as reported in the paper ``The decay modes of proton drip-line nuclei with A between 42 and 47'' \cite{1992Bor01}. A 69 A$\cdot$MeV $^{58}$Ni beam was incident on a natural nickel target and the projectile fragments were separated using the Ligne d'Ions Super Epluch\'{e}s (LISE) spectrometer. The isotopes were identified by time of flight and energy loss measurements. ``$^{43}$Cr is identified here for the first time, with 264 events recorded.'' The half-life of $^{43}$Cr was determined via maximum-likelihood analysis of the time spectrum to be 21$^{+4}_{-3}$ms, which agrees with the presently accepted value of 21.6(7)~ms.

\subsection{$^{44}$Cr}\vspace{0.0cm}

The 1987 paper ``Direct Observation of New Proton Rich Nuclei in the Region 23$\leq$Z$\leq$29 Using A 55A$\cdot$MeV $^{58}$Ni Beam'' reported the first observation of $^{44}$Cr at GANIL by Pougheon et al. \cite{1987Pou01}. The fragmentation of a 55 A$\cdot$MeV $^{58}$Ni beam on a nickel target was used to produce proton-rich isotopes which were separated with the LISE spectrometer. Energy loss, time of flight, and magnetic rigidity measurements were recorded. ``Here $^{44}$Cr (T$_Z = -2$) is unambiguously identified with a statistics of 9 counts.''

\subsection{$^{45}$Cr}\vspace{0.0cm}
``A New Delayed Proton Precursor: Chromium-45'' was published in 1974 by Jackson et al. reporting the discovery of $^{45}$Cr \cite{1974Jac01}. The isotope was produced by the $^{32}$S($^{16}$O,3n) fusion evaporation reaction with $^{16}$O beams of energies between 50 and 82 MeV from the Chalk River MP tandem accelerator. Spectra of $\beta$-delayed protons were measured with a surface barrier counter telescope. ``The new T$_z = -3/2$ isotope, $^{45}$Cr, has been produced by the $^{32}$S($^{16}$O,3n)$^{45}$Cr reaction. Its half-life was measured to be 50$\pm$6~ms.'' This half-life is consistent with the presently accepted value of 60.9(4)~ms.

\subsection{$^{46}$Cr}\vspace{0.0cm}
Zioni et al. published the first observation of $^{46}$Cr in the paper ``An Investigation of Proton-Rich Nuclei and Other Products from the Bombardment of $^{24}$Mg, $^{28}$Si and $^{32}$S by $^{16}$O Ions'' in 1972 \cite{1972Zio01}. At the Jerusalem Racah Institute a $^{16}$O beam was accelerated to 22$-$33 MeV with an EN tandem and $^{46}$Cr was produced in the fusion-evaporation $^{32}$S($^{16}$O,2n) reaction on a zinc sulphide target. Beta- and $\gamma$-ray spectra were recorded with a NE102 plastic scintillator and a Ge(Li) detector, respectively. ``In particular the mass excess of the previously unobserved nucleus $^{46}$Cr is found to be $-$29.46$\pm$0.03~MeV; its half-life is 0.26$\pm$0.6~s.'' This half-life corresponds to the currently accepted value. A previously reported half-life of 1.1~s \cite{1954Tyr01} was evidently incorrect.

\subsection{$^{47}$Cr}\vspace{0.0cm}
In the paper entitled ``New Proton-Rich Nuclei in the f${_{7/2}}$ Shell'', Proctor et al. described the discovery of $^{47}$Cr in 1972 \cite{1972Pro01}. The Michigan State University sector-focused cyclotron accelerated $^{3}$He to 70.8~MeV and the reaction $^{50}$Cr($^{3}$He,$^{6}$He) was used to produce $^{47}$Cr. The outgoing $^6$He particles were detected in the focal plane of an Enge split-pole magnetic spectrograph. ``The present measurements represent the first observation of $^{47}$Cr, $^{51}$Fe, and $^{55}$Ni.'' Previously, a half-life of 430~ms \cite{1954Tyr01} was assigned to either $^{47}$Cr or $^{49}$Mn. Similarly a 400~ms half-life \cite{1954Tyr01} was assigned to either $^{47}$Cr or $^{46}$V.

\subsection{$^{48}$Cr}\vspace{0.0cm}
Rudstam et al. reported on the discovery of $^{48}$Cr in their 1952 publication ``Nuclear Reactions of Iron with 340-Mev Protons'' \cite{1952Rud01}. Protons were accelerated to 340 MeV by the Berkeley 184 inch cyclotron and $^{48}$Cr was produced in spallation reactions on iron targets. Decay curves were measured with a chlorine-quenched Amperex Geiger-M\"uller tube following chemical separation. ``In the chromium decay curves a new activity was found after subtraction of the activity due to 26.5-day Cr$^{51}$... The chromium isotope from these experiments can be assigned the mass number 48.'' In three different runs half-lives of 19, 24, and 23~h were measured which agree with the currently accepted value of 21.56(3)~h.

\subsection{$^{49}$Cr}\vspace{0.0cm}
``Artificial Radioactivity of $^{49}$Cr'', published in 1942 by O'Connor et al., announced the discovery of $^{49}$Cr \cite{1942OCo01}. The bombardment of TiO$_2$ targets with 20 MeV alpha particles from the Ohio State cyclotron resulted in formation of $^{49}$Cr by the reaction $^{46}$Ti($\alpha$,n). A Wulf quartz fiber electrometer connected to a Freon filled ionization chamber was used to measure decay and absorption curves. ``Since chemical separation shows that the 41.9-minute activity, produced by alpha-particle bombardment of titanium and by fast neutron bombardment of chromium is an isotope of chromium and since this period has not been found by proton bombardment of vanadium or deuteron bombardment of chromium, the activity must evidently be due to $^{49}$Cr.'' This 41.9(3)~min half-life agrees with the accepted half-life measurement of 42.3(1)~min.

\subsection{$^{50}$Cr}\vspace{0.0cm}
The discovery of stable $^{50}$Cr was reported in the 1930 paper ``Constitution of Chromium'' by Aston \cite{1930Ast03}. $^{50}$Cr was observed with the Cavendish mass spectrometer using the volatile compound Cr(CO)$_6$. ``The intensity of the beam of mass-rays has been so increased that not only has it been possible, by the fine slits, to obtain a value for the packing fraction of Cr$^{52}$ but also, by the use of coarse slits and long exposures, to reveal no less than three new isotopes, and to determine their relative abundances...'' The three new isotopes were $^{50}$Cr, $^{53}$Cr and $^{54}$Cr.

\subsection{$^{51}$Cr}\vspace{0.0cm}
In the paper ``K-Electron Capture and Internal Conversion in Cr$^{51}$'', Walke et al. described their discovery of $^{51}$Cr in 1940 \cite{1940Wal03}. At Berkeley, a sample of metallic titanium was bombarded with 16-MeV alpha particles and $^{51}$Cr was produced in the reaction $^{48}$Ti($\alpha$,n). Electrons, X-rays and $\gamma$-rays were recorded.  ``As a result of these experiments we have failed to observe the 600-day vanadium activity which is consistent with the previous assignment to V$^{47}$. However, we have discovered another isotope which decays by K-electron capture, and we ascribe it to Cr$^{51}$.'' Its half life was measured to be 26.5(1)~d, which is consistent with the accepted value of 27.7010(11)~d.

\subsection{$^{52}$Cr}\vspace{0.0cm}
Aston discovered stable $^{52}$Cr in 1923 as reported in the paper ``Further Determinations of the Constitution of the Elements by the Method of Accelerated Anode Rays'' \cite{1923Ast01}. No details regarding the mass spectroscopic observation of cobalt is given. ``Vanadium and chromium give single mass-lines at positions expected from their atomic weights 51 and 52.''

\subsection{$^{53,54}$Cr}\vspace{0.0cm}
The discovery of the stable isotopes $^{53}$Cr and $^{54}$Cr was reported in the 1930 paper ``Constitution of Chromium'' by Aston \cite{1930Ast03}. $^{50}$Cr was observed with the Cavendish mass spectrometer using the volatile compound Cr(CO)$_6$. ``The intensity of the beam of mass-rays has been so increased that not only has it been possible, by the fine slits, to obtain a value for the packing fraction of Cr$^{52}$ but also, by the use of coarse slits and long exposures, to reveal no less than three new isotopes, and to determine their relative abundances...'' The three new isotopes were $^{50}$Cr, $^{53}$Cr and $^{54}$Cr.

\subsection{$^{55}$Cr}\vspace{0.0cm}
In the 1952 publication ``$^{55}$Cr, ein neues Chrom-Isotop mit T = 3,52 min Halbwertszeit'' Flammersfeld and Herr reported the discovery of $^{55}$Cr \cite{1952Fla01}. Slow neutrons produced in the reaction of deuterium on beryllium at Mainz University bombarded a pure chromium-oxide target and $^{55}$Cr was produced by neutron capture on $^{54}$Cr. Decay curves were measured following chemical separation. ``Es lag nahe, die gemessene Aktivit\"at auf ein durch unvermeidbare schnelle Neutronen aus dem Hauptisotop des Chroms nach $^{52}$Cr(n,p) gebildetes $^{52}$V (T = 3,77 min) zur\"uckzuf\"uhren, um so mehr als Absorptionsversuche nahezu die gleiche $\beta$-Energie wie beim $^{52}$V ergaben. Bestrahlungen mit und ohne Cadmium zeigten aber, da\ss\ die Aktivit\"at aus dem Chrom haupts\"achlich mit thermischen Neutronen entsteht, also ein (n,$\gamma$)-Proze\ss\ vorliegen mu\ss.'' [The measured activity could have been from the decay of $^{52}$V (T = 3.77 min) produced in the reaction $^{52}$Cr(n,p) due to the presence of fast neutrons, especially because absorption measurement resulted in $\beta$-decay energies similar to the decay of $^{52}$V. However, irradiations with and without cadmium demonstrated that the activity was produced predominantly by thermal neutrons, thus it was due to the (n,$\gamma$) process.] The measured half-life of 3.52(3)~min agrees with the currently accepted value of 3.497(3)~min. Previously reported half-lives of 1.7~h \cite{1937Poo01}, 2.27~h \cite{1940Dic01}, 1.6~h \cite{1940Ama01}, and 1.3~h \cite{1947Ser01} were evidently incorrect.

\subsection{$^{56}$Cr}\vspace{0.0cm}
In 1960, Dropesky et al. published the discovery of $^{56}$Cr in their paper ``Note on the decay of the new nucleide Cr$^{56}$'' \cite{1960Dro01}. Normal chromium metal was bombarded with 2.7$-$2.9 MeV tritons at Los Alamos National Laboratory, and $^{56}$Cr was formed in the reaction $^{54}$Cr(t,p). The emission of $\beta$- and $\gamma$-rays was measured following chemical separation. ``First evidence for the presence of Cr$^{56}$ in the purified Cr sample came from the growth of 2.6-h Mn$^{56}$ activity as identified by the 0.845, 1.81, and 2.13 MeV photopeaks observed with a 3$\times$3 NaI(Tl) scintillator.'' The half-life was measured to be 5.94(1)~min, which is still the currently accepted half-life. Previous searches for $^{56}$Cr which was predicted to have a substantially longer half-live were unsuccessful \cite{1957Jon01,1957Roy03}.

\subsection{$^{57}$Cr}\vspace{0.0cm}
Davids et al. reported the discovery of $^{57}$Cr in the 1978 paper entitled ``Mass and $\beta$ decay of the new isotope $^{57}$Cr'' \cite{1978Dav01}. A 21 MeV $^{11}$B beam from the Argonne FN tandem accelerator bombarded an enriched $^{48}$Ca target and $^{57}$Cr was created with the fusion evaporation reaction $^{48}$Ca($^{11}$B,pn). An NE102-plastic scintillator was used to measure $\beta$-rays and $\gamma$-rays were detected with Ge(Li) detectors. ``$^{57}$Cr was identified by the observation of a decaying 205.8-keV $\gamma$ ray in the singles spectra... The half-life of $^{57}$Cr was obtained from the decay of the 205.8-keV $\gamma$ ray, and is 21.1$\pm$1.0~s.'' This half-life is still the presently accepted value.

\subsection{$^{58,59}$Cr}\vspace{0.0cm}
Guerreau et al. reported the discovery of $^{58}$Cr, $^{59}$Cr in the 1980 paper ``Seven New Neutron Rich Nuclides Observed in Deep Inelastic Collisions of 340 MeV $^{40}$Ar on $^{238}$U'' \cite{1980Gue01}. A 340 MeV $^{40}$Ar beam accelerated by the Orsay ALICE accelerator facility bombarded a 1.2 mg/cm$^2$ thick UF$_4$ target supported by an aluminum foil. The isotopes were identified using two $\Delta$E-E telescopes and two time of flight measurements. ``The new nuclides $^{54}$Ti, $^{56}$V, $^{58-59}$Cr, $^{61}$Mn, $^{63-64}$Fe, have been produced through $^{40}$Ar + $^{238}$U reactions.'' At least twenty counts were recorded for these isotopes.

\subsection{$^{60}$Cr}\vspace{0.0cm}
The first tentative identification of $^{60}$Cr was reported in 1980 by Breuer et al., and the new isotope was reported in their paper ``Production of neutron-excess nuclei in $^{56}$Fe-induced reactions'' \cite{1980Bre01}. A beam of $^{56}$Fe ions with an energy of 8.3 MeV/u from the Berkeley SuperHILAC bombarded $^{209}$Bi and $^{238}$U targets and products of this reaction were identified with a $\Delta$E-E time-of-flight detector placed 118 cm from the target. ``In addition, tentative identification of six additional nuclides ($^{56}$Ti, $^{57-58}$V, $^{60}$Cr, $^{61}$Mn, and $^{63}$Fe) is reported.'' The definite discovery of $^{60}$Cr was announced eight years later by Bosch et al. in ``Beta- and gamma-decay studies of neutron-rich chromium, manganese, cobalt and nickel isotopes including the new isotopes $^{60}$Cr and $^{60g}$Mn'' \cite{1988Bos01} by measuring a half-life of 0.57(6)~s which is consistent with the currently adopted value of 0.49(1)~s.

\subsection{$^{61,62}$Cr}\vspace{0.0cm}
The 1985 paper ``Production and Identification of New Neutron-Rich Fragments from 33 MeV/u $^{86}$Kr Beam in the 18$\leq$Z$\leq$27 Region'' by Guillemaud-Mueller et al. reported the first observation of $^{61}$Cr and $^{62}$Cr \cite{1985Gui01}. The 33 MeV/u $^{86}$Kr beam bombarded tantalum targets and the fragments were separated with the GANIL triple-focusing analyser LISE. ``Each particle is identified by an event-by-event analysis. The mass A is determined from the total energy and the time of flight, and Z by the $\Delta$E and E measurements... In addition to that are identified the following new isotopes: $^{47}$Ar, $^{57}$Ti, $^{59,60}$V, $^{61,62}$Cr, $^{65,65}$Mn, $^{66,67,68}$Fe, $^{68,69,70}$Co.''

\subsection{$^{63,64}$Cr}\vspace{0.0cm}
In their paper ``New neutron-rich isotopes in the scandium-to-nickel region, produced by fragmentation of a 500 MeV/u $^{86}$Kr beam'', Weber et al. presented the first observation of $^{63}$Cr and $^{64}$Cr in 1992 \cite{1992Web01}. The isotopes were produced in the fragmentation reaction of a 500 A$\cdot$MeV $^{86}$Kr beam from the heavy-ion synchrotron SIS on a beryllium target and separated with the zero-degree spectrometer FRS at GSI. ``The isotope identification was based on combining the values of B$\rho$, time of flight (TOF), and energy loss ($\triangle$E) that were measured for each ion passing through the FRS and its associated detector array... The results shown in [the figure] represent unambiguous evidence for the production of the very neutron-rich isotopes $^{58}$Ti, $^{61}$V, $^{63}$Cr, $^{66}$Mn, $^{69}$Fe, and $^{71}$Co, and yield indicative evidence for the production of $^{64}$Cr, $^{72}$Co, and $^{75}$Ni.'' Thirty-five and three counts of $^{63}$Cr and $^{64}$Cr were recorded, respectively. The discovery of $^{64}$Cr was confirmed by Sorlin et al. in 1999 \cite{1999Sor01,2000Sor01}.

\subsection{$^{65-67}$Cr}\vspace{0.0cm}
Bernas et al. observed $^{65}$Cr, $^{66}$Cr, and $^{67}$Cr for the first time in 1997 as reported in their paper ``Discovery and cross-section measurement of 58 new fission products in projectile-fission of 750$\cdot$A MeV $^{238}$U'' \cite{1997Ber01}. Uranium ions were accelerated to 750 A$\cdot$MeV by the GSI UNILAC/SIS accelerator facility and bombarded a beryllium target. The isotopes produced in the projectile-fission reaction were separated using the fragment separator FRS and the nuclear charge Z for each was determined by the energy loss measurement in an ionization chamber. ``The mass identification was carried out by measuring the time of flight (TOF) and the magnetic rigidity B$\rho$ with an accuracy of 10$^{-4}$.''  82, 19, and 4 counts of $^{65}$Cr, $^{66}$Cr and $^{67}$Cr were observed, respectively.

\subsection{$^{68}$Cr}\vspace{0.0cm}
$^{68}$Cr was discovered by Tarasov et al. in 2009 and published in ``Evidence for a change in the nuclear mass surface with the discovery of the most neutron-rich nuclei with 17 $\le$ Z $\le$ 25'' \cite{2009Tar01}. $^9$Be targets were bombarded with 132 MeV/u $^{76}$Ge ions accelerated by the Coupled Cyclotron Facility at the National Superconducting Cyclotron Laboratory at Michigan State University. $^{68}$Cr was produced in projectile fragmentation reactions and identified with a two-stage separator consisting of the A1900 fragment separator and the S800 analysis beam line. ``The observed fragments include fifteen new isotopes that are the most neutron-rich nuclides of the elements chlorine to manganese ($^{50}$Cl, $^{53}$Ar, $^{55,56}$K, $^{57,58}$Ca, $^{59,60,61}$Sc, $^{62,63}$Ti, $^{65,66}$V, $^{68}$Cr, $^{70}$Mn).''

\section{Discovery of $^{46-70}$Mn}

Twentyfive manganese isotopes from A = $46-70$ have been discovered so far; these include 1 stable, 9 proton-rich and 15 neutron-rich isotopes.  According to the HFB-14 model \cite{2007Gor01}, $^{78}$Mn should be the last odd-odd particle stable neutron-rich nucleus while the odd-even particle stable neutron-rich nuclei should continue through $^{87}$Mn. The proton dripline has been reached and no more long-lived isotopes are expected to exist because $^{44}$Mn and $^{45}$Mn have been shown to be unbound with upper lifetime limits of 105~ns and 70~ns, respectively \cite{1992Bor01}. About 13 isotopes have yet to be discovered corresponding to 34\% of all possible manganese isotopes.

\begin{figure}
	\centering
	\includegraphics[scale=.5]{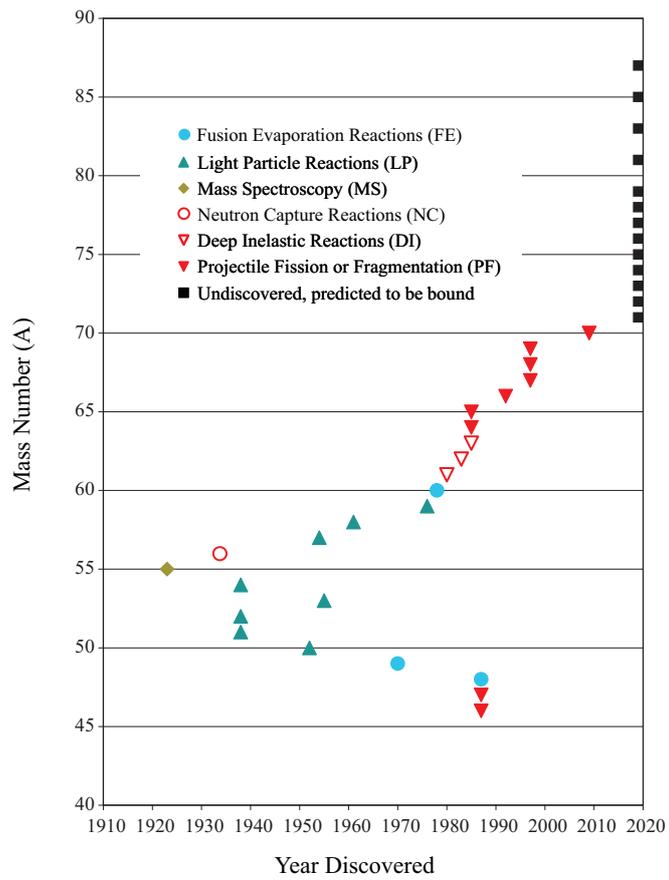}
	\caption{Manganese isotopes as a function of time when they were discovered. The different production methods are indicated. The solid black squares on the right hand side of the plot are isotopes predicted to be bound by the HFB-14 model.}
\label{f:year-mn}
\end{figure}

Figure \ref{f:year-mn} summarizes the year of first discovery for all manganese isotopes identified by the method of discovery. The range of isotopes predicted to exist is indicated on the right side of the figure. The radioactive manganese isotopes were produced using deep-inelastic reactions (DI), heavy-ion fusion-evaporation (FE), light-particle reactions (LP), neutron-capture reactions (NC), and projectile fragmentation of fission (PF). The stable isotope was identified using mass spectroscopy (MS). Heavy ions are all nuclei with an atomic mass larger than A=4 \cite{1977Gru01}. Light particles also include neutrons produced by accelerators. In the following, the discovery of each manganese isotope is discussed in detail and a summary is presented in Table 1.

\subsection{$^{46,47}$Mn}\vspace{0.0cm}

The 1987 paper ``Direct Observation of New Proton Rich Nuclei in the Region 23$\leq$Z$\leq$29 Using A 55A$\cdot$MeV $^{58}$Ni Beam'', reported the first observation of $^{46}$Mn and $^{47}$Mn by Pougheon et al. \cite{1987Pou01}. The fragmentation of a 55 A$\cdot$MeV $^{58}$Ni beam at GANIL on a nickel target was used to produce proton-rich isotopes which were separated with the LISE spectrometer. Energy loss, time of flight, and magnetic rigidity measurements were made such that ``two additional Mn isotopes were identified: $^{47}$Mn and $^{46}$Mn with respectively 335 and 15 counts.''

\subsection{$^{48}$Mn}\vspace{0.0cm}

$^{48}$Mn was discovered in 1987 by Sekine et al. as described in the paper ``The Beta Decay of $^{48}$Mn: Gammow-Teller quenching in fp-shell nuclei'' \cite{1987Sek01}. At the UNILAC at GSI, Darmstadt an 11.4 MeV/u $^{40}$Ca beam was aimed at a FEBIAD-F ion source with a graphite catcher and $^{48}$Mn was produced in the fusion evaporation reaction $^{12}$C($^{40}$Ca,p3n). The isotopes were separated by mass in a magnetic sector field and identified by $\gamma$- and $\beta$-ray decay spectroscopy. ``After accounting for single and double escape peaks as well as for peaks due to summing with 511 keV annihilation radiation, and leaving the $\gamma$-lines at 1660 and 2167 keV unassigned, 16 $\gamma$-transitions were ascribed to the decay of $^{48}$Mn.'' A half-life of 150(10)~ms was recorded, which is in agreement with the currently accepted value of 158.1(22)~ms.

\subsection{$^{49}$Mn}\vspace{0.0cm}

In 1970, Cerny et al. published the paper ``Heavy-Ion Reactions as a Technique for Direct Mass Measurements of Unknown Z$>$N Nuclei'' \cite{1970Cer02} describing the discovery of the isotope $^{49}$Mn. A 27.5~MeV $^{12}$C beam was directed at calcium targets at the Oxford EN tandem Van De Graaff so that the products of the reaction could be detected in a counter telescope made of four semiconductor detectors. ``As an initial experiment, we have chosen the reaction $^{40}$Ca($^{12}$C,t)$^{49}$Mn as a means of studying the hitherto uncharacterized nuclide $^{49}$Mn; our results agree well with the theoretical prediction of its mass.'' A half-life of 0.43~s had been previously been reported \cite{1954Tyr01} but this half-life could not be definitively attributed to either $^{49}$Mn or $^{47}$Cr.

\subsection{$^{50}$Mn}\vspace{0.0cm}

Martin and Breckon discovered $^{50}$Mn in 1952 which they announced in ``The New Radioactive Isotopes Vanadium 46, Manganese 50, Cobalt 54'' \cite{1952Mar01}. Protons with energies between 15 and 22 MeV from the McGill University cyclotron bombarded titanium targets and $^{50}$Mn was produced in the reaction $^{50}$Cr(p,n)$^{50}$Mn. Positron activities were displayed on a cathode-ray oscilloscope and photographs of the screen were taken for subsequent graphical analysis. The assignment of $^{50}$Mn was based on the threshold energy and the \textit{ft} value. ``One is thus led to assign the 0.40, 0.28, and 0.18 sec. activities to the isotopes V$^{46}$, Mn$^{50}$, and Co$^{54}$, respectively.'' The measured half-life agrees with the presently accepted value of 283.88(46)~ms.

\subsection{$^{51,52}$Mn}\vspace{0.0cm}

In 1938 Livingood and Seaborg outlined their discovery of $^{51}$Mn and $^{52}$Mn in the article ``Radioactive Manganese Isotopes'' \cite{1938Liv02}. The reactions for these isotopes made use of deuterons at energies of 5.5 and 7.6 MeV, and helium ions at energies of 16 MeV at the Berkeley Cyclotron. The bombardment of $^{50}$Cr by deuterons and neutrons was used to yield $^{51}$Mn, while the reaction of $^{54}$Fe with deuterons and alpha particles was chosen for $^{52}$Mn. Decay curves were measured with a quartz fiber electroscope following chemical separation. ``The Cr(d,n)Mn reaction could lead to Mn$^{51}$, Mn$^{53}$, Mn$^{54}$, and of these possibilities we believe the 46-minute activity must be assigned to Mn$^{51}$... Two positron emitting manganese isotopes Mn$^{52}$ and Mn$^{54}$ can be expected through the disintegration type Fe(d,$\alpha$)Mn; nevertheless, we believe both these activities must be described as isomers of Mn$^{52}$.'' The extracted half-lives of 46(2)~min ($^{51}$Mn) and 21(2)~min and 6.5(1)~d for $^{52}$Mn agree with the accepted values of 46.2(1)~min, 21.1(2)~min and 5.591(3)~d for $^{51}$Mn and $^{52}$Mn, respectively. The 21~min half-life corresponds to an isomer of $^{52}$Mn. Half-lives of 5~d \cite{1937Liv01}, 21~min \cite{1937Liv01,1937Dar01} and 42~min \cite{1938Dub01} had previously been reported, however, no mass assignments were made.

\subsection{$^{53}$Mn}\vspace{0.0cm}

Wilkinson and Sheline reported the discovery of $^{53}$Mn in their 1955 paper ``New Isotope of Manganese-53'' \cite{1955Wil01}. A sample of enriched $^{53}$Cr was bombarded with 9.5 MeV protons from the Berkeley 60-inch cyclotron. $^{53}$Mn was produced in the $^{53}$Cr(p,n) charge-exchange reaction and its activity was measured using a Geiger counter following chemical separation. ``It is established then that we have a long-lived gammaless orbital electron capturing isotope of manganese. In view of the isotopic composition of the enriched Cr$^{53}$ isotope used in the proton bombardment, the presence of an unassigned position in the nuclear periodic table at 25 protons and 28 neutrons, the unusual stability expected because of shell closure at 28 neutrons, and finally, the fact that this is a proton excess nucleus since it is an orbital electron capturer, the only reasonable assignment of this long-lived manganese activity is Mn$^{53}$." The half-life was not directly measured, only calculated from comparing the cross sections with the charge exchange reaction on $^{54}$Cr. A half-life of 140 years was deduced assuming a 5/2 ground state for $^{53}$Mn. The possibility of a 7/2 ground state was considered which would result in a half-life of $\sim$10$^{6}$~y. The latter turned out to be correct with an accepted half-life for $^{53}$Mn of 3.74(4)$\times$10$^{6}$~y.

\subsection{$^{54}$Mn}\vspace{0.0cm}

In 1938 Livingood and Seaborg outlined their discovery of $^{54}$Mn in the article ``Radioactive Manganese Isotopes'' \cite{1938Liv02}. The reaction for this isotope made use of deuterons at energies of 5.5 and 7.6 MeV, and helium ions at energies of 16 MeV from the Berkeley Cyclotron. $^{54}$Mn was produced through the activation of iron with deuterons, chromium with deuterons, and vanadium with helium ions. ``Only two choices are available for this isotope when produced from iron by deuteron bombardment; Mn$^{52}$ and Mn$^{54}$. We have given evidence above for the assignment of Mn$^{52}$ of the isomeric activities with 21 minutes and 6 days half-lives, so that the 310-day period is to be associated with Mn$^{54}$.'' The measured half-life of 310(20)~d is consistent with the accepted value of 312.05(4)~d.

\subsection{$^{55}$Mn}\vspace{0.0cm}

In 1923 Aston stated the discovery of the only stable manganese isotope, $^{55}$Mn in ``Further Determinations of the Constitution of the Elements by the Method of Accelerated Anode Rays'' \cite{1923Ast01}. No details regarding the mass spectroscopic observation of manganese was given. ``Manganese behaved surprisingly well, and yielded unequivocal results indicating that it is a simple element of mass number 55.''

\subsection{$^{56}$Mn}\vspace{0.0cm}

In 1934 Amaldi et al. discovered $^{56}$Mn which was announced in ``Radioactivity Produced by Neutron Bombardment-V'' \cite{1934Ama01}. Neutrons from beryllium powder mixed with emanation (radon) irradiated manganese targets at the Istituto Fisica della R. Universit\`a in Rome, Italy. The $\beta$-ray activity was measured with a Geiger-M\"{u}ller counter. ``For manganese, besides the period of 4 minutes, another of about 150 minutes was observed, the active substance of which cannot be separated from manganese and is probably Mn$^{56}$, which also is obtained from iron and cobalt.'' The half-life mentioned above of 150~min agrees well with the accepted value of 2.5789(1)~h.

\subsection{$^{57}$Mn}\vspace{0.0cm}

In the 1954 paper titled ``Manganese-57" \cite{1954Coh01}, Cohen et al. published the discovery of the isotope $^{57}$Mn. Isotopically enriched $^{57}$Fe was bombarded with neutrons produced by the Oak Ridge 86 inch cyclotron and $^{56}$Mn was created by the (n,p) charge exchange reaction. The $\gamma$-ray spectrum was examined with a NaI(T1) scintillation spectrometer, while the $\beta$-ray spectrum was analyzed with an anthracene scintillation spectrometer. The isotope ``was identified by chemical separations, by measurement of its formation and cross section, by investigation of possible impurity effects, and by comparison of its gamma spectrum with that of $^{57}$Co." The recorded half-life of 1.7(1)~min is in agreement with the accepted value of 85.4(18)~s. A previous attempt to observe $^{57}$Mn was not successful \cite{1950Nel01}.

\subsection{$^{58}$Mn}\vspace{0.0cm}

The discovery of $^{58}$Mn by Chittenden et al. is outlined in the 1961 paper ``New Isotope of Manganese; Cross Sections of the Iron Isotopes for 14.8-MeV Neutrons'' \cite{1961Chi01}. At the University of Arkansas 400-kV Cockcroft-Walton accelerator, enriched iron was bombarded by 14.8 MeV neutrons and the $\gamma$-ray spectrum was analyzed by a NaI(T1) scintillation spectrometer. ``On the basis of cross-bombardments utilizing chemical separations and gamma-ray spectrum (which fits the measured energy level of Fe$^{58}$) the activity is assigned to Mn$^{58}$ from the Fe$^{58}$(n,p) and Fe$^{57}$(n,np) reactions.'' The isotope was assigned a half-life of 1.1(1)~min, which corresponds to an isomer of $^{58}$Mn with a currently accepted half-life of 65.2(5)~s.

\subsection{$^{59}$Mn}\vspace{0.0cm}

In 1976, Kashy et al. published the discovery of $^{59}$Mn in their paper ``Observation of highly neutron-rich $^{43}$Cl and $^{59}$Mn'' \cite{1976Kas01}. An enriched $^{64}$Ni foil was bombarded with a 74 MeV $^{3}$He beam at the Michigan State University Cyclotron and $^{59}$Mn was produced in the reaction $^{64}$Ni($^3$He,$^8$B). The momentum of the ejectiles was examined with an Enge split pole spectrograph, energy loss was measured, and time of flight data was taken from a plastic scintillator to verify the observations and record the mass of the isotope. ``[The figure] shows spectra of $^8$B ions from the $^{64}$Ni and $^{27}$Al targets,... A peak with a width of about 200 keV is observed for the $^{59}$Mn.''

\subsection{$^{60}$Mn}\vspace{0.0cm}

Norman et al. presented their 1978 discovery of $^{60}$Mn in the paper ``Mass and $\beta$-decay of the new neutron-rich isotope $^{60}$Mn'' \cite{1978Nor01}. An enriched $^{48}$Ca foil was bombarded with $^{18}$O ions at an energy of 56 MeV at the Argonne tandem accelerator. $^{60}$Mn was produced in the fusion-evaporation reaction $^{48}$Ca($^{16}$O, $\alpha$pn). Gamma-rays were analyzed with Ge(Li) detectors, and $\beta$-rays were examined using plastic scintillators. ``Gamma rays have been attributed to the decay of $^{60}$Mn by comparison of their energies with those of previously reported levels of $^{60}$Fe and by their coincidence relations with known gamma rays of $^{60}$Fe. From these measurements, the half life, ground-state spin, parity, and decay scheme of $^{60}$Mn were determined.'' The half-life was measured to be 1.79(1)~s, which corresponds to an isomer of $^{60}$Mn with a half-life of 1.77(2)~s.

\subsection{$^{61}$Mn}\vspace{0.0cm}

Guerreau et al. reported the discovery of $^{61}$Mn in the 1980 paper ``Seven New Neutron Rich Nuclides Observed in Deep Inelastic Collisions of 340 MeV $^{40}$Ar on $^{238}$U'' \cite{1980Gue01}. A 340 MeV $^{40}$Ar beam accelerated by the Orsay ALICE accelerator facility bombarded a 1.2 mg/cm$^2$ thick UF$_4$ target supported by an aluminum foil. $^{61}$Mn was identified using two $\Delta$E-E telescopes and two time of flight measurements. ``The new nuclides $^{54}$Ti, $^{56}$V, $^{58-59}$Cr, $^{61}$Mn, $^{63-64}$Fe, have been produced through $^{40}$Ar + $^{238}$U reactions.'' At least twenty counts were recorded for these isotopes.

\subsection{$^{62}$Mn}\vspace{0.0cm}

Runte et al. reported the discovery of $^{62}$Mn in 1983: ``Decay Studies of Neutron-Rich Products from $^{76}$Ge Induced Multinucleon Transfer Reactions Including the New Isotopes $^{62}$Mn, $^{63}$Fe, and $^{71,72,73}$Cu'' \cite{1983Run01}. A 9 MeV/u $^{76}$Ge beam from the GSI UNILAC accelerator was used to bombard a natural W target and $^{62}$Mn was produced in deep inelastic reactions. The reaction products were collected in a graphite catcher inside a FEBIAD-E ion source and separated with an on-line mass separator. ``In addition to the known nuclides of the elements chromium through germanium with A=56-75, we identified the decays of $^{62}$Mn, $^{63}$Fe, and $^{71,72,73}$Cu, of which only $^{63}$Fe was known from direct particle-identification measurements to be bound." The reported half-life of 0.88(15)~s corresponds to the currently adopted value. Three years earlier only hints for the existence of $^{62}$Mn were reported \cite{1980Gue01}.

\subsection{$^{63}$Mn}\vspace{0.0cm}

Bosch et al. discovered $^{63}$Mn in 1985 as described in ``Beta-decay half-lives of new neutron-rich chromium-to-nickel isotopes and their consequences for the astrophysical r-process'' \cite{1985Bos01}. $^{76}$Ge was accelerated to 11.4 MeV/u at GSI and bombarded a natural tungsten target. $^{63}$Mn was produced in a multinucleon transfer reaction and separated with the FEBIAD-F ion source and the GSI on-line mass separator. ``Beta-decay studies of the new neutron-rich isotopes $^{58,59}$Cr, $^{63}$Mn, $^{66,67}$Co and $^{69}$Ni, yielding distinctly shorter half-lives than the corresponding theoretical predictions, are presented.'' The measured half-life of 0.25(4)~s for $^{63}$Mn agrees with the currently accepted value of 0.275(4)~s.

\subsection{$^{64,65}$Mn}\vspace{0.0cm}

The 1985 paper ``Production and Identification of New Neutron-Rich Fragments from 33 MeV/u $^{86}$Kr Beam in the 18$\leq$Z$\leq$27 Region'' by Guillemaud-Mueller et al. reported the first observation of $^{64}$Mn and $^{65}$Mn \cite{1985Gui01}. The 33 MeV/u $^{86}$Kr beam bombarded tantalum targets and the fragments were separated with the GANIL triple-focusing analyser LISE. ``Each particle is identified by an event-by-event analysis. The mass A is determined from the total energy and the time of flight, and Z by the $\Delta$E and E measurements... In addition to that are identified the following new isotopes $^{47}$Ar, $^{57}$Ti, $^{59,60}$V, $^{61,62}$Cr, $^{65,65}$Mn, $^{66,67,68}$Fe, $^{68,69,70}$Co.''

\subsection{$^{66}$Mn}\vspace{0.0cm}

In their paper ``New neutron-rich isotopes in the scandium-to-nickel region, produced by fragmentation of a 500 MeV/u $^{86}$Kr beam'', Weber et al. presented the first observation of $^{66}$Mn in 1992 \cite{1992Web01}. $^{66}$Mn was produced in the fragmentation reaction of a 500 A$\cdot$MeV $^{86}$Kr beam from the heavy-ion synchrotron SIS on a beryllium target and separated with the zero-degree spectrometer FRS at GSI. ``The isotope identification was based on combining the values of B$\rho$, time of flight (TOF), and energy loss ($\triangle$E) that were measured for each ion passing through the FRS and its associated detector array.'' Sixteen counts of $^{66}$Mn were recorded. The previously reported ``...hints for the observation of $^{54}$Sc and $^{66}$Mn'' \cite{1985Gui01} was not considered to be sufficient to warrant discovery.

\subsection{$^{67-69}$Mn}\vspace{0.0cm}

Bernas et al. observed $^{67}$Mn, $^{68}$Mn, and $^{69}$Mn for the first time in 1997 as reported in their paper ``Discovery and cross-section measurement of 58 new fission products in projectile-fission of 750$\cdot$A MeV $^{238}$U'' \cite{1997Ber01}. Uranium ions were accelerated to 750 A$\cdot$MeV by the GSI UNILAC/SIS accelerator facility and bombarded a beryllium target. The isotopes produced in the projectile-fission reaction were separated using the fragment separator FRS and the nuclear charge Z for each was determined by the energy loss measurement in an ionization chamber. ``The mass identification was carried out by measuring the time of flight (TOF) and the magnetic rigidity B$\rho$ with an accuracy of 10$^{-4}$.'' 245, 43 and 5 counts of $^{67}$Mn, $^{68}$Mn and $^{69}$Mn were observed, respectively.

\subsection{$^{70}$Mn}\vspace{0.0cm}

$^{70}$Mn was discovered by Tarasov et al. in 2009 and published in ``Evidence for a change in the nuclear mass surface with the discovery of the most neutron-rich nuclei with 17 $\le$ Z $\le$ 25'' \cite{2009Tar01}. $^9$Be targets were bombarded with 132 MeV/u $^{76}$Ge ions accelerated by the Coupled Cyclotron Facility at the National Superconducting Cyclotron Laboratory at Michigan State University. $^{70}$Mn was produced in projectile fragmentation reactions and identified with a two-stage separator consisting of the A1900 fragment separator and the S800 analysis beam line. ``The observed fragments include fifteen new isotopes that are the most neutron-rich nuclides of the elements chlorine to manganese ($^{50}$Cl, $^{53}$Ar, $^{55,56}$K, $^{57,58}$Ca, $^{59,60,61}$Sc, $^{62,63}$Ti, $^{65,66}$V, $^{68}$Cr, $^{70}$Mn).''

\section{Discovery of $^{48-78}$Ni}

Thirty-one nickel isotopes from A = $48-78$ have been discovered so far; these include 5 stable, 11 proton-rich and 15 neutron-rich isotopes.  According to the HFB-14 model \cite{2007Gor01}, $^{87}$Ni should be the last odd-even particle stable neutron-rich nucleus while the even-even particle stable neutron-rich nuclei should continue through $^{94}$Ni. At the proton dripline $^{47}$Ni could still be particle stable. Thus, about 14 isotopes have yet to be discovered corresponding to about 31\% of all possible nickel isotopes.

\begin{figure}
	\centering
	\includegraphics[scale=.5]{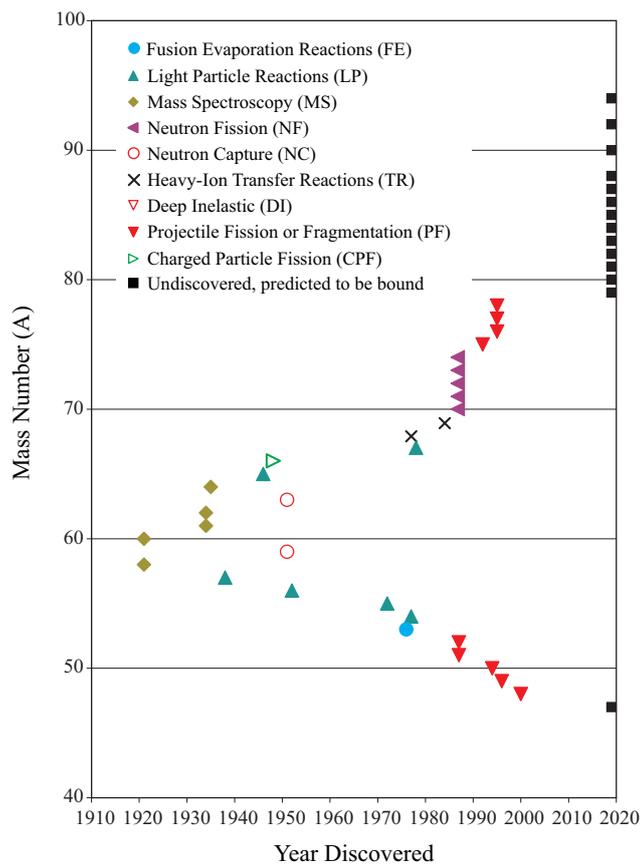}
	\caption{Nickel isotopes as a function of time when they were discovered. The different production methods are indicated. The solid black squares on the right hand side of the plot are isotopes predicted to be bound by the HFB-14 model.}
\label{f:year-ni}
\end{figure}

Figure \ref{f:year-ni} summarizes the year of first discovery for all nickel isotopes identified by the method of discovery. The range of isotopes predicted to exist is indicated on the right side of the figure. The radioactive nickel isotopes were produced using deep-inelastic reactions (DI), heavy-ion fusion-evaporation (FE), light-particle reactions (LP), neutron-induced fission (NF), charged-particle induced fission (CPF), neutron-capture reactions (NC), heavy-ion transfer (TR), and projectile fragmentation of fission (PF). The stable isotope was identified using mass spectroscopy (MS). Heavy ions are all nuclei with an atomic mass larger than A=4 \cite{1977Gru01}. In the following, the discovery of each nickel isotope is discussed in detail and a summary is presented in Table 1.

\subsection{$^{48}$Ni}\vspace{0.0cm}
In the paper ``Discovery of Doubly Magic $^{48}$Ni'', Blank et al. reported the discovery of $^{48}$Ni in 2000 \cite{2000Bla01}. A natural nickel target was bombarded by a 74.5 MeV/nucleon beam of $^{58}$Ni from the GANIL cyclotrons. $^{48}$Ni was separated and identified with the SISS/LISE3 facility. ``Because of the efficiency of the MCP detector part of the statistics is lost in this spectrum. Nevertheless, two events of $^{48}$Ni are clearly observed... In this spectrum, we observe four counts which can be unambiguously attributed to $^{48}$Ni.''

\subsection{$^{49}$Ni}\vspace{0.0cm}
In the paper ``First Observation of the T$_{z}=-$7/2 Nuclei $^{45}$Fe and $^{49}$Ni,'' Blank et al. reported the discovery of $^{49}$Ni in 1996 \cite{1996Bla01}. A 600 A$\cdot$MeV $^{58}$Ni beam from the SIS synchrotron bombarded a beryllium target and isotopes were separated with the projectile-fragment separator FRS. $^{49}$Ni was identified by time-of-flight, $\Delta$E, and B$\rho$ analysis. ``We observed ten events of $^{42}$Cr, three events of $^{45}$Fe, and five events of $^{49}$Ni. These three isotopes have been identified for the first time in the present experiment.''

\subsection{$^{50}$Ni}\vspace{0.0cm}
The 1994 paper ``Production cross sections and the particle stability of proton-rich nuclei from $^{58}$Ni fragmentation'' reported the discovery of $^{50}$Ni by Blank et al. \cite{1994Bla01}. A 650 MeV/nucleon beam of $^{58}$Ni from the SIS synchrotron bombarded a beryllium target and $^{50}$Ni was separated and identified using the FRS separator. ``For the yet-unobserved isotope $^{50}$Ni, we find three counts which fulfill the conditions we impose on the TOF, on the energy loss, and on the position at the final focus.''

\subsection{$^{51,52}$Ni}\vspace{0.0cm}
The 1987 paper ``Direct Observation of New Proton Rich Nuclei in the Region 23$\leq$Z$\leq$29 Using A 55A$\cdot$MeV $^{58}$Ni Beam,'' reported the first observation of $^{51}$Ni and $^{52}$Ni by Pougheon et al. \cite{1987Pou01}. The fragmentation of a 55 A$\cdot$MeV $^{58}$Ni beam at GANIL on nickel and aluminum targets was used to produce proton-rich isotopes which were separated with the LISE spectrometer. Energy loss, time of flight, and magnetic rigidity measurements were made. ``Here, $^{52}$Ni (T$_z=-$2) and $^{51}$Ni (T$_z=-$5/2) are identified with respectively 68 and 7 counts.''

\subsection{$^{53}$Ni}\vspace{0.0cm}
Vieira et al. reported the observation of excited states of $^{53}$Ni in 1976 in the paper ``Extension of the T$_{z}=-$3/2 Beta-Delayed Proton Precursor Series to $^{57}$Zn'' \cite{1976Vie01}. The Berkeley 88-in. cyclotron accelerated $^{16}$O beams to 60 and 65 MeV which then bombarded calcium targets and $^{53}$Ni was produced in the fusion-evaporation reaction $^{40}$Ca($^{16}$O,3n). Beta-delayed protons were measured with a semiconductor counter telescope. ``The most reasonable source of the 1.94 MeV activity is $\beta$-delayed proton emission following the decay of $^{53}$Ni produced via the $^{40}$Ca($^{16}$O,3n) reaction.'' The measured half-life of 45(15)~ms corresponds to the currently accepted value.

\subsection{$^{54}$Ni}\vspace{0.0cm}
In the paper ``Mass measurements of the proton-rich nuclei $^{50}$Fe and $^{54}$Ni,'' Tribble et al. reported the discovery of $^{54}$Ni in 1977 \cite{1977Tri02}. Alpha particles accelerated to 110 MeV with the Texas A\&M University 88-inch Cyclotron were used to produce the reaction $^{58}$Ni($^{4}$He,$^{8}$He) and the ejectiles were observed at the focal plane of an Enge split-pole magnetic spectrograph. ``The experiments provide the first observation and subsequent mass measurement of the proton-rich nuclei $^{50}$Fe and $^{54}$Ni.'' The measured $\beta$-decay energy was 7.77(5)~MeV which was used to estimate a half-life of 140~ms; this is close to the adopted value of 104(7)~ms.

\subsection{$^{55}$Ni}\vspace{0.0cm}
In the paper entitled ``New Proton-Rich Nuclei in the f${_{7/2}}$ Shell,'' Proctor et al. described the discovery of $^{55}$Ni in 1972 \cite{1972Pro01}. The Michigan State University sector-focused cyclotron accelerated $^{3}$He to 65$-$75~MeV and the reaction $^{58}$Ni($^{3}$He,$^{6}$He) was used to produce $^{55}$Ni. The outgoing $^6$He particles were detected in the focal plane of an Enge split-pole magnetic spectrograph. ``The present measurements represent the first observation of $^{47}$Cr, $^{51}$Fe, and $^{55}$Ni.''

\subsection{$^{56}$Ni}\vspace{0.0cm}
In his 1952 paper ``Nickel 56,'' Worthington reported on the discovery of $^{56}$Ni \cite{1952Wor01}. At Berkeley, a 340~MeV proton beam irradiated zinc foils and $^{56}$Ni was identified by measuring the decay with a Geiger counter following chemical separation. ``From this work it may be deduced that (1) the half-life of Ni$^{56}$ should be 6.0$\pm$0.5 days; (2) there are at least four gamma-rays associated with the disintegration, with energies of approximately 0.16, 0.5, 0.8, and $>$1.4 Mev; (3) the decay is mainly by electron capture rather than positron emission.'' The measured half-life agrees with the presently adopted value of 6.075(10)~d. An independent observation \cite{1952She01} of $^{56}$Ni was submitted only three weeks after the submission by Worthington. It should be mentioned that Aston had reported tentative evidence that $^{56}$Ni would be stable \cite{1934Ast01}.

\subsection{$^{57}$Ni}\vspace{0.0cm}
The discovery of $^{57}$Ni was reported in the 1938 paper ``Radio Isotopes of Nickel'' by Livingood and Seaborg \cite{1938Liv04}. $^{57}$Ni was observed at the University of California, Berkeley, by irradiating iron with 12.6 and 16 MeV $\alpha$-particles. Positrons and $\gamma$-rays were measured following chemical separation. ``We wish to report a new radioactive isotope of nickel, formed as the result of the exposure of iron to several microampere hours of bombardment with helium ions at 12.6 Mev and also at 16 Mev... We have not been able to detect this activity after strong irradiation of nickel with deuterons or slow neutrons, so we feel justified in ascribing  the activity to Ni$^{57}$ through Fe$^{54}$($\alpha$,n)Ni$^{57}$.'' The reported 36(2)~h half-life agrees with the currently accepted value of 35.60(6)~h.

\subsection{$^{58}$Ni}\vspace{0.0cm}
In the paper ``The Constitution of Nickel,'' Aston described the discovery of stable $^{58}$Ni in 1921 \cite{1921Ast02}. At the Cavendish Laboratory in Cambridge, England a discharge tube with a mixture of nickel carbonyl vapor and carbon dioxide was used to obtain mass spectra. ``The spectrum consists of two lines, the stronger at 58 and the weaker at 60... Nickel therefore consists of at least two isotopes.''

\subsection{$^{59}$Ni}\vspace{0.0cm}
In 1959, Brosi et al. identified $^{59}$Ni in 1951 as reported in the paper ``Characteristics of Ni$^{59}$ and Ni$^{63}$'' \cite{1951Bro01}. Enriched $^{58}$Ni targets were irradiated at the Oak Ridge reactor and activities were measured with Geiger-M\"uller counters following chemical separation. ``The data used in the calculation are given in [the table] along with estimated probable errors. These data give 7.5$\pm$1.3$\times$10$^4$ yrs for the K electron capture half-life of Ni$^{59}$.'' This half-life agrees with the currently accepted value of 7.6$\times$10$^4$~y.

\subsection{$^{60}$Ni}\vspace{0.0cm}
In the paper ``The Constitution of Nickel,'' Aston described the discovery of stable $^{60}$Ni in 1921 \cite{1921Ast02}. At the Cavendish Laboratory in Cambridge, England a discharge tube with a mixture of nickel carbonyl vapor and carbon dioxide was used to obtain mass spectra. ``The spectrum consists of two lines, the stronger at 58 and the weaker at 60... Nickel therefore consists of at least two isotopes.''

\subsection{$^{61,62}$Ni}\vspace{0.0cm}
The discovery of the stable isotopes $^{61}$Ni and $^{62}$Ni was published in the 1934 paper ``Constitution of Carbon, Nickel, and Cadmium'' by Aston \cite{1934Ast01}. The isotopes were identified with the Cavendish Laboratory mass spectrograph. ``The analysis of nickel by means of its carbonyl has been repeated, and the more intense mass-spectra obtained reveal two new isotopes 62 and 61.''

\subsection{$^{63}$Ni}\vspace{0.0cm}
Brosi et al. identified $^{63}$Ni in 1951 as reported in the paper ``Characteristics of Ni$^{59}$ and Ni$^{63}$'' \cite{1951Bro01}. Enriched $^{62}$Ni targets were irradiated at the Oak Ridge reactor and activities were measured with Geiger-M\"uller counters following chemical separation. ``The half-life of 85 yrs for Ni$^{63}$ calculated from the data in [the table] is considerably shorter than values previously estimated from activation data.'' The measured half-life of 85(20)~y is close to the presently adopted value of 101.2(15)~y. Fourteen years earlier, an activity of 160(10)~min was assigned to either $^{63}$Ni or $^{65}$Ni \cite{1937Oes01}. In addition, activities of a few hours \cite{1935Rot01}, 2.5~h \cite{1937Hey01}, 2.60(3)~h \cite{1938Liv04} and 2.6~h \cite{1942Nel01} were incorrectly assigned to $^{63}$Ni.

\subsection{$^{64}$Ni}\vspace{0.0cm}
The existence of stable $^{64}$Ni was demonstrated by deGier and Zeeman at the University of Amsterdam in 1935 and reported in the paper ``The Isotopes of Nickel'' \cite{1935deG02}. Volatile carbonyl of nickel was used in a discharge tube in front of a mass spectrograph. ``The very first photos were already interesting. Line 56 was absent and 64 was clearly visible... Line 64 was easy to get and appeared to have the same relative intensity during the whole experiment. This indicated clearly that it was not due to an adventitious compound.'' In 1934 Aston was not confident in the observation of $^{64}$Ni: ``The analysis of nickel by means of its carbonyl has been repeated, and the more intense mass-spectra obtained reveal two new isotopes 62 and 61. Lines at 56 and 64 present to less than 1 per cent are probably due to isotopes, but this is not yet certain.''

\subsection{$^{65}$Ni}\vspace{0.0cm}
Swartout et al. reported the discovery of $^{65}$Ni in the 1946 paper ``Mass Assignment of 2.6~h Ni$^{65}$'' \cite{1946Swa01}.
Enriched $^{63}$Cu and $^{65}$Cu were irradiated with neutrons from the Oak Ridge uranium pile. $^{65}$Ni was formed in the (n,p) charge-exchange reaction and $\beta$- and $\gamma$-rays were measured following chemical separation. ``The availability of enriched copper isotopes in the Manhattan Project has now made possible a positive assignment of the 2.6~h Ni isotope to a mass number of 65.'' This half-life agrees with the currently accepted value of 2.5172(3)~h. Nine years earlier, an activity of 160(10)~min was assigned to either $^{63}$Ni or $^{65}$Ni \cite{1937Oes01}.

\subsection{$^{66}$Ni}\vspace{0.0cm}
The discovery of $^{66}$Ni was reported by Goeckermann and Perlman in the 1948 publication ``Characteristics of Bismuth Fission with High Energy Particles'' \cite{1948Goe01}. 200 MeV deuterons from the 184-in Berkeley cyclotron were used to produce fragments from the fission of bismuth. ``The following is a list of bismuth fission products which were identified in the present studies (references are given for isotopes which only recently appeared in the literature or which have been hitherto unreported): Ca$^{45}$, Fe$^{59}$, Ni$^{66}$, Cu$^{67}$..." The half-life of $^{66}$Ni was given in a footnote: ``A 56-hr. $\beta^-$-emitter which proved to be the parent of 5-min Cu$^{66}$.'' This half-life agrees with the presently accepted value of 54.6(3)~h.

\subsection{$^{67}$Ni}\vspace{0.0cm}
Kouzes et al. identified $^{67}$Ni in the 1978 paper `` Mass of $^{67}$Ni'' \cite{1978Kou02}. An enriched $^{70}$Zn target was bombarded with a 56 MeV $^4$He beam from the Princeton University AVF cyclotron. $^{67}$Ni was produced in the multiple particle transfer reaction $^{70}$Zn($^4$He,$^7$Be)$^{67}$Ni, separated with the quadrupole-dipole-dipole-dipole spectrograph, and measured with two resistive-wire gas-proportional counters and a plastic scintillator. ``This measurement gives a result for the Q value of $-$19164$\pm$22 keV measured relative to the Q value of $-$18512$\pm$2 keV for $^{25}$Mg($^4$He,$^7$Be)$^{22}$Ne. Using $-$69560$\pm$3 keV for the $^{70}$Zn mass excess this gives a $^{67}$Ni mass excess of $-$63741$\pm$22 keV.'' Previous observations of half-lives of 50(3)~s \cite{1965Mea01}, 18(4)~s \cite{1971Taf01}, and 16(4)~s \cite{1975Rei01} were evidently incorrect. The latter two assignments were based on $\gamma$-ray measurements which were reassigned by Runte et al. to $^{70}$Cu.

\subsection{$^{68}$Ni}\vspace{0.0cm}
In their paper ``Masses of $^{62}$Fe and the New Isotope $^{68}$Ni from ($^{18}$O,$^{20}$Ne) Reactions,'' Bhatia et al. presented the first observation of $^{68}$Ni in 1977 \cite{1977Bha01}. At the Heidelberg MP tandem beams of $^{18}$O with energies of 81-84 MeV were incident on enriched $^{70}$Zn targets. Reaction products were detected by a $\Delta$E-E counter and analyzed by a Q3D Spectrograph. ``The utility of the ($^{18}$O,$^{20}$Ne) reaction at small angles for the mass determination of highly neutron rich nuclei is demonstrated by a determination of the mass excess of $^{62}$Fe as ($-$58.946$\pm$0.022)~MeV and of $^{68}$Ni as ($-$63.466$\pm$0.028)~MeV.''

\subsection{$^{69}$Ni}\vspace{0.0cm}
Dessagne et al. observed $^{69}$Ni for the first time in 1984 as reported in their paper ``The Complex Transfer Reaction ($^{14}$C,$^{15}$O) on Ni, Zn, and Ge Targets: Existence and Mass of $^{69}$Ni'' \cite{1984Des01}. A 72 MeV beam of $^{14}$C provided by the Orsay MP tandem was incident on an enriched $^{70}$Zn target. $^{69}$Ni was identified by measuring the $^{15}$O ejectiles in the double-focussing n=1/2 magnetic spectrometer BACCHUS. ``In the case of the $^{68}$Zn and $^{70}$Zn targets, the ($^{14}$C,$^{15}$O) reactions clearly populate the ground states of $^{67}$Ni and $^{69}$Ni... The $^{69}$Ni nucleus is observed and its mass measured here for the first time.''

\subsection{$^{70-74}$Ni}\vspace{0.0cm}
In the 1987 paper ``Identification of the New-Neutron Rich Isotopes $^{70-74}$Ni and $^{74-77}$Cu in Thermal Neutron Fission of $^{235}$U'' Armbruster et al. described the discovery of $^{70}$Ni, $^{71}$Ni, $^{72}$Ni, $^{73}$Ni, and $^{74}$Ni \cite{1987Arm01}. At the Institut Laue-Langevin in Grenoble, France, fragments from thermal neutron induced fission of $^{235}$U were analyzed in the LOHENGRIN recoil separator. ``Events associated with different elements are well enough separated to show unambiguously the occurrence of the isotopes $^{70-74}$Ni and $^{74-77}$Cu among other isotopes already known."

\subsection{$^{75}$Ni}\vspace{0.0cm}
In their paper ``New neutron-rich isotopes in the scandium-to-nickel region, produced by fragmentation of a 500 MeV/u $^{86}$Kr beam,'' Weber et al. presented the first observation of $^{75}$Ni in 1992 \cite{1992Web01}. $^{75}$Ni was produced in the fragmentation reaction of a 500 A$\cdot$MeV $^{86}$Kr beam from the heavy-ion synchrotron SIS on a beryllium target and separated with the zero-degree spectrometer FRS at GSI. ``The results...represent unambiguous evidence for the production of the very neutron-rich isotopes $^{58}$Ti, $^{61}$V, $^{63}$Cr, $^{66}$Mn, $^{69}$Fe, and $^{71}$Co, and yield indicative evidence for the production of $^{64}$Cr, $^{72}$Co, and $^{75}$Ni.'' Two counts of $^{75}$Ni were observed. The discovery of $^{75}$Ni was confirmed by Ameil et al. in 1998 \cite{1998Ame01}.

\subsection{$^{76-78}$Ni}\vspace{0.0cm}
In 1995 Engelmann et al. reported the discovery of $^{76}$Ni, $^{77}$Ni, and $^{78}$Ni, in ``Production and Identification of Heavy Ni Isotopes: Evidence for the Doubly Magic Nucleus $^{78}_{28}$Ni'' \cite{1995Eng01}. $^{238}$U ions were accelerated in the UNILAC and the heavy-ion synchrotron SIS at GSI to an energy of 750 A-MeV. The nickel isotopes were produced by projectile fission, separated in-flight by the FRS and identified event-by-event by measuring magnetic rigidity, energy loss and time of flight. ``For a total dose of 10$^{13}$ U ions delivered in 132 h on the target three events can be assigned to the isotope $^{78}$Ni. Other new nuclei, $^{77}$Ni, $^{73,74,75}$Co and $^{80}$Cu can be identified, the low count rate requires a background-free measurement.'' One hundred thirty two, 13, and 3 counts were observed for $^{76}$Ni, $^{77}$Ni and $^{78}$Ni, respectively. $^{76}$Ni was not considered a new nucleus quoting an internal report \cite{1995Ame01}. The results of this report were subsequently published in a conference proceeding \cite{1995Ame02}, however, publication in a refereed journal occurred only three years later \cite{1998Ame01}.

\section{Discovery of $^{55-80}$Cu}

Twentysix copper isotopes from A = $55-80$ have been discovered so far; these include 2 stable, 9 proton-rich and 15 neutron-rich isotopes.  According to the HFB-14 model \cite{2007Gor01}, $^{92}$Cu should be the last odd-odd particle stable neutron-rich nucleus while the odd-even particle stable neutron-rich nuclei should continue through $^{103}$Cu. $^{55}$Cu is the most proton-rich long-lived copper isotope because $^{54}$Cu has been shown to be unbound with an upper half-life limit of 75~ns \cite{1994Bla01}. About 18 isotopes have yet to be discovered corresponding to 41\% of all possible copper isotopes.

\begin{figure}
	\centering
	\includegraphics[scale=.5]{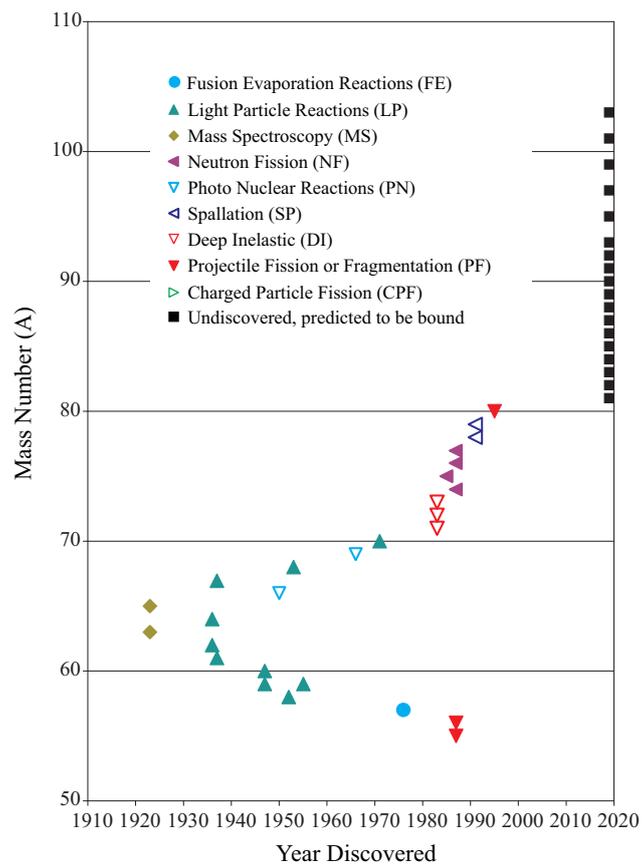}
	\caption{Copper isotopes as a function of time when they were discovered. The different production methods are indicated. The solid black squares on the right hand side of the plot are isotopes predicted to be bound by the HFB-14 model.}
\label{f:year-co}
\end{figure}

Figure \ref{f:year-co} summarizes the year of first discovery for all copper isotopes identified by the method of discovery. The range of isotopes predicted to exist is indicated on the right side of the figure. The radioactive copper isotopes were produced using deep-inelastic reactions (DI), heavy-ion fusion-evaporation (FE), light-particle reactions (LP), neutron-induced fission (NF), charged-particle induced fission (CPF), photo nuclear reactions (PN), spallation (SP), and projectile fragmentation of fission (PF). The stable isotope was identified using mass spectroscopy (MS). Heavy ions are all nuclei with an atomic mass larger than A=4 \cite{1977Gru01}. Light particles also include neutrons produced by accelerators. In the following, the discovery of each copper isotope is discussed in detail and a summary is presented in Table 1.

\subsection{$^{55,56}$Cu}\vspace{0.0cm}

The 1987 paper ``Direct Observation of New Proton Rich Nuclei in the Region 23$\leq$Z$\leq$29 Using A 55A$\cdot$MeV $^{58}$Ni Beam'', reported the first observation of $^{55}$Cu and $^{56}$Cu by Pougheon et al. \cite{1987Pou01}. The fragmentation of a 55 A$\cdot$MeV $^{58}$Ni beam at GANIL on nickel and aluminum targets was used to produce proton-rich isotopes which were separated with the LISE spectrometer. Energy loss, time of flight, and magnetic rigidity measurements were made. ``The spectra show evidence for $^{56}$Cu and $^{55}$Cu (1420 and 75 events respectively).''

\subsection{$^{57}$Cu}\vspace{0.0cm}

Vieira et al. reported the observation of excited states of $^{57}$Cu in 1976 in the paper ``Extension of the T$_{z}=-$3/2 Beta-Delayed Proton Precursor Series to $^{57}$Zn'' \cite{1976Vie01}. The Berkeley 88-in. cyclotron accelerated $^{20}$Ne beams to 62 and 70 MeV which then bombarded calcium targets and $^{57}$Zn was produced in the fusion-evaporation reaction $^{40}$Ca($^{20}$Ne,3n). $^{57}$Cu was populated by $\beta$-decay and delayed protons were measured with a semiconducting counter telescope. ``The groups observed at 4.65~MeV and 1.95~MeV can be assigned to the isospin-forbidden proton decay of the lowest T=3/2 state of $^{57}$Cu to the ground state and the first excited state of $^{56}$Ni, respectively.'' The ground state of $^{57}$Cu was discovered eight years later by Shinozuka et al. \cite{1984Shi02}.

\subsection{$^{58}$Cu}\vspace{0.0cm}

$^{58}$Cu was measured by Martin and Breckon in 1952 as described in ``The New Radioactive Isotopes Vanadium 46, Manganese 50, Cobalt 54'' \cite{1952Mar01}. Nickel foils were bombarded with 15 MeV protons from the McGill cyclotron and $^{58}$Cu was formed in (p,n) charge-exchange reactions. Activation measurements were performed with a pneumatic target extractor and a scintillation counter. ``Analysis of the photographs gave the mean values of 0.873 sec. and 3.04 sec. respectively for the half lives of Sc$^{41}$ and Cu$^{58}$.'' This value for the half-life agrees with the presently accepted value of 3.204(7)~s. The authors did not consider their measurement of $^{58}$Cu a new discovery citing a data compilation \cite{1950NBS01}. However, this compilation only referred to a private communication. Previous reports of half-lives of 80(2)~s \cite{1938Rid01}, 81(2)~s \cite{1939Del01}, and 10~min \cite{1947Lei01} were evidently incorrect. The first two half-lives most likely corresponded to $^{59}$Cu.

\subsection{$^{59,60}$Cu}\vspace{0.0cm}

Leith et al. reported the first observation of $^{59}$Cu and $^{60}$Cu in the 1947 paper ``Radioactivity of Cu$^{60}$'' \cite{1947Lei01}. Protons from 5 to 15 MeV from the Berkeley 37-in. frequency-modulated cyclotron bombarded separated $^{58}$Ni and $^{60}$Ni targets. Activities were measured with an ionization chamber and Ryerson-Lindemann electrometer. They tentatively assigned $^{59}$Cu: ``The 81-second and 7.9-minute positron activities produced by proton bombardment of Ni, observed by Delsasso, et al., and tentatively assigned to either Cu$^{58}$ or Cu$^{60}$, correspond to 81-second and 10-minute activities after bombarding Ni$^{58}$ with protons in the 37-in. cyclotron. These are tentatively assigned to Cu$^{59}$ and Cu$^{58}$, respectively, on the basis of threshold and excitation considerations.'' This 81~s half-life for $^{59}$Cu agrees with the currently accepted value of 81.5(5)~s, however, the half-life for $^{58}$Cu was not verified. In 1955, Lindner et al. described the definite discovery of $^{59}$Cu in ``Radiations of copper 59'' \cite{1955Lin02}. $^{59}$Cu was identified by bombarding nickel foils with 22 MeV deuterons measuring a half-life of 81(1)~s.

The assignment of $^{60}$Cu was firm: ``Chemical separation of normal nickel targets after bombardment with 15-Mev and 6-Mev protons into Cu, Ni, and Co fractions, accomplished within one hour, showed in each case that more than 99 percent of the 24.6-minute activity followed the Cu-separation chemistry. Mass separation in a calutron accomplished within one hour of the proton bombardment, showed without question that this activity belonged to Cu$^{60}$.'' This half-life is part of the weighted average that comprises the currently accepted value of 23.7(4)~min. The previous tentative assignment of 7.9~min to $^{60}$Cu \cite{1938Rid01,1939Del01} was evidently incorrect.

\subsection{$^{61}$Cu}\vspace{0.0cm}

In 1937 Ridenour and Henderson discovered $^{61}$Cu, which they outlined in their paper ``Artificial Radioactivity Produced by Alpha-Particles'' \cite{1937Rid01}. Alpha particles accelerated to 9~MeV by the Princeton cyclotron bombarded nickel targets and $^{61}$Cu was produced in the reaction $^{58}$Ni($\alpha$,p). The positron emissions were measured through their absorption in aluminum and the element assignment was achieved by chemical separation. ``The half life of Cu$^{61}$ is 3.4$\pm$0.1 hours; both the half-life and the upper limit of the beta-ray spectrum agree with the values determined by Thornton for the same radioelement obtained in the bombardment of Ni with deuterons.'' This half-life agrees with the presently accepted value of 3.333(5)~h. In the quoted paper by Thornton, no mass assignment for the measured half-life was made \cite{1937Tho01}.

\subsection{$^{62}$Cu}\vspace{0.0cm}

Heyn published his discovery of $^{62}$Cu in the 1936 paper ``Evidence for the Expulsion of Two Neutrons from Copper and Zinc by One Fast Neutron'' \cite{1936Hey01}. Fast neutrons produced in the Li + $^{2}$H reaction bombarded a copper target at the X-Ray Research Laboratory in Eindhoven, Netherlands. Decay curves were measured following chemical separation. ``An investigation of the particles emitted by copper (10.5-minute period) by means of magnetic deflection in vacuo and a Geiger-M\"uller counter proved these to be positrons... From these data we may infer that the following reactions take place with fast neutrons: $^{63}$Cu + $^1$n $\rightarrow ^{62}$Cu + 2 $^1$n, $^{62}$Cu $\rightarrow ^{62}$Ni + e$^+$.'' The measured half-life of 10.5(5)~min is consistent with the current value of 9.673(8)~min. A previous report of a stable $^{62}$Cu isotope \cite{1923Dem01} was evidently incorrect.

\subsection{$^{63}$Cu}\vspace{0.0cm}

In the paper entitled ``The Mass-spectrum of Copper" Aston published the discovery of stable $^{63}$Cu in 1923 \cite{1923Ast02}. Mass spectra were taken with cuprous chloride in the accelerated anode ray method. ``The lines are faint, but their evidence is conclusive since they appear at the expected positions 63 and 65 and have the intensity ratio, about 2.5 to 1, predicted from the chemical atomic weight 63.57.''

\subsection{$^{64}$Cu}\vspace{0.0cm}

Van Voorhis reported the discovery of $^{64}$Cu in the 1936 paper ``The Artificial Radioactivity of Copper, a Branch Reaction'' \cite{1936Voo01}. Copper targets were bombarded with 5 to 6~MeV deuterons accelerated with the Berkeley ``magnetic resonance accelerator or cyclotron.'' The activities were measured with a pressure ionization chamber and FP-54 Pliotron. The upper boundaries of energy were studied, as well as absorption curves for the positron and electron activities that were analyzed from data taken in an ionization chamber. ``...the half-life of both positron and electron activities was found to be exactly the same, a more exact measurement giving the value 12.8$\pm$0.1 hours.'' This half-life measurement is in accordance with the accepted value of 12.701(2)~h. A previous report of a stable $^{64}$Cu isotope \cite{1923Dem01} was evidently incorrect.

\subsection{$^{65}$Cu}\vspace{0.0cm}

In the paper entitled ``The Mass-spectrum of Copper'' Aston published the discovery of stable $^{65}$Cu in 1923 \cite{1923Ast02}. Mass spectra were taken with cuprous chloride in the accelerated anode ray method. ``The lines are faint, but their evidence is conclusive since they appear at the expected positions 63 and 65 and have the intensity ratio, about 2.5 to 1, predicted from the chemical atomic weight 63.57.''

\subsection{$^{66}$Cu}\vspace{0.0cm}

In 1937 Chang et al. described the observation of $^{66}$Cu in ``Radioactivity Produced by Gamma Rays and Neutrons of High Energy'' \cite{1937Cha01}. Deuterons accelerated by a voltage of 520~kV were used to produce neutrons: ``In gallium bombarded with neutrons from lithium + deuterons and boron + deuterons a new radioactivity decaying with a half-period of about 5~min and having an intensity similar to that of the 60 min period has been found. As this period seems to be identical with one of the periods produced in copper by neutron capture, it is probably due to $^{66}$Cu, formed according to the reaction: $^{66}_{31}$Ga + $^1_0$n $\rightarrow ^{66}_{29}$Cu + $^4_2$He.''  This half-life agrees with the currently accepted value of 5.120(14)~min. The neutron capture measurement on copper mentioned in the quote refers to a paper by Amaldi et al. who had reported the 5~min half-life without a mass assignment \cite{1935Ama01}. In 1936 Van Voorhis had suggested that this half-life would correspond to $^{66}$Cu based on his measurement of $^{64}$Cu. A previous report of a stable $^{66}$Cu isotope \cite{1923Dem01} was evidently incorrect.

\subsection{$^{67}$Cu}\vspace{0.0cm}

The discovery of $^{67}$Cu was reported by Goeckermann and Perlman in the 1948 publication ``Characteristics of Bismuth Fission with High Energy Particles'' \cite{1948Goe01}. 200 MeV deuterons from the 184-in Berkeley cyclotron were used to produce fragments from the fission of bismuth. ``The following is a list of bismuth fission products which were identified in the present studies: Ca$^{45}$, Fe$^{59}$, Ni$^{66}$, Cu$^{67}$..." The half-life of $^{67}$Cu was given in a footnote: ``A 56-hr. $\beta^-$-emitter tentatively assigned to Cu$^{67}$ on the basis of the mode of disintegration and half-life.'' This half-life is close to the currently adopted value of 61.83(12)~h. Less than a month later Hopkins et al. published the observation of $^{67}$Cu independently \cite{1948Hop01}.

\subsection{$^{68}$Cu}\vspace{0.0cm}

$^{68}$Cu was observed by Flammersfeld in 1953 and reported in ``$^{68}$Cu, ein neues Kupfer-Isotop mit T = 32 sec Halbwertszeit'' \cite{1953Fla01}. Fast neutrons produced by the bombardment of 1.4~MeV deuterons on lithium irradiated zinc targets and the activity was measured with 100 $\mu$-counters. ``Bei der Bestrahlung von Zink mit energiereichen Neutronen (Li + D-Neutronen, E$_D$ = 1.4~MeV) tritt eine neue Halbwertszeit von T = 32$\pm$2 sec auf...'' [The irradiation of zinc with energetic neutrons (Li + D-neutrons, E$_D$ = 1.4~MeV) results in a new half-life of T = 32$\pm$2~s...] This half-life is included in the current average value of 31.1(15)~s.

\subsection{$^{69}$Cu}\vspace{0.0cm}

Van Klinken et al. published the discovery of $^{69}$Cu in the 1966 paper ``Decay of a New Isotope: Cu$^{69}$'' \cite{1966Kli01}. Bremsstrahlung from the 70-MeV Iowa State synchrotron irradiated enriched $^{70}$Zn targets. Beta- and gamma-ray spectra were recorded to identify $^{69}$Cu produced in photo-nuclear reactions. ``The half-life of Cu$^{69}$ was first measured by following the decay rate for different parts of the beta-ray spectrum and for prominent lines in the NaI(Tl) gamma-ray spectrum. Our estimate from these data is T$_{1/2}$ = 2.8$\pm$0.3 min.'' This half-life is part of the weighted average for the accepted value of 2.85(15)~min.

\subsection{$^{70}$Cu}\vspace{0.0cm}

In 1971 $^{70}$Cu was first observed by Taff et al. in ``The Decays of $^{70a}$Cu, $^{70b}$Cu and $^{67}$Ni'' \cite{1971Taf01}. An enriched $^{70}$Zn metal bead was bombarded with 14-MeV neutrons produced in the reaction $^3$He(d,n)$^4$He from the Oak Ridge Cockcroft-Walton cascade generator. $^{70}$Cu was produced in the (n,p) charge exchange reaction and identified by $\beta$- and $\gamma$-rays measured with plastic, Na(Tl) scintillators and Ge(Li) detectors. ``Three activities with half-lives of 5$\pm$1~s, 42$\pm$3~s and 18$\pm$4~s, have been assigned to two isomers of $^{70}$Cu [from $^{70}$Zn(n,p)], and to $^{67}$Ni [from $^{70}$Zn(n,$\alpha$)], respectively.'' The 5$\pm$1~s half-life corresponds to an isomer and the 42$\pm$3~s half-life is consistent with the presently accepted value for the ground state of 44.5(2)~s.

\subsection{$^{71-73}$Cu}\vspace{0.0cm}

Runte et al. reported the discovery of $^{71}$Cu, $^{72}$Cu, and $^{73}$Cu in 1983: ``Decay Studies of Neutron-Rich Products from $^{76}$Ge Induced Multinucleon Transfer Reactions Including the New Isotopes $^{62}$Mn, $^{63}$Fe, and $^{71,72,73}$Cu'' \cite{1983Run01}. A 9 MeV/u $^{76}$Ge beam from the GSI UNILAC accelerator was used to bombard a natural tungsten target and the copper isotopes were produced in deep inelastic reactions. The reaction products were collected in a graphite catcher inside a FEBIAD-E ion source and separated with an on-line mass separator.  ``In addition to the known nuclides of the elements chromium through germanium with A=56-75, we identified the decays of $^{62}$Mn, $^{63}$Fe, and $^{71,72,73}$Cu, of which only $^{63}$Fe was known from direct particle-identification measurements to be bound." The following half-lives were obtained for $^{71-73}$Cu, respectively: 19.5(16), 6.6(1) and 3.9(3) s. The value for $^{71}$Cu is the currently accepted half-life. The half-life measurement for $^{72}$Cu is in agreement with the accepted value of 6.63(3)~s. The half-life given for $^{73}$Cu is used in the calculation of the weighted average of the currently accepted value of 4.2(3)~s.

\subsection{$^{74}$Cu}\vspace{0.0cm}

In the 1987 paper ``Identification of the New-Neutron Rich Isotopes $^{70-74}$Ni and $^{74-77}$Cu in Thermal Neutron Fission of $^{235}$U'' Armbruster et al. described the discovery of $^{74}$Cu \cite{1987Arm01}. At the Institut Laue-Langevin in Grenoble, France, fragments from thermal neutron induced fission of $^{235}$U were analyzed in the LOHENGRIN recoil separator. ``Events associated with different elements are well enough separated to show unambiguously the occurrence of the isotopes $^{70-74}$Ni and $^{74-77}$Cu among other isotopes already known."

\subsection{$^{75}$Cu}\vspace{0.0cm}

Reeder et al. announced the discovery of $^{75}$Cu in 1985 as described in ``Delayed neutron precursor $^{75}$Cu'' \cite{1985Ree01}. Fragments from neutron-induced fission of $^{235}$U were extracted from a FEBIAD ion source at Brookhaven and $^{75}$Cu was separated and identified with the TRISTAN on-line isotope separator facility. ``A new delayed neutron precursor with a half-life of 1.3$\pm$0.1~s has been observed at mass 75... Mass formula and fission yield predictions indicate that $^{75}$Cu is the most likely precursor." This half-life is consistent with the presently adopted value of 1.224(3)~s.

\subsection{$^{76,77}$Cu}\vspace{0.0cm}

In the 1987 paper ``Identification of the New-Neutron Rich Isotopes $^{70-74}$Ni and $^{74-77}$Cu in Thermal Neutron Fission of $^{235}$U'' Armbruster et al. described the discovery of $^{76}$Cu and $^{77}$Cu \cite{1987Arm01}. At the Institut Laue-Langevin in Grenoble, France, fragments from thermal neutron induced fission of $^{235}$U were analyzed in the LOHENGRIN recoil separator. ``Events associated with different elements are well enough separated to show unambiguously the occurrence of the isotopes $^{70-74}$Ni and $^{74-77}$Cu among other isotopes already known."

\subsection{$^{78,79}$Cu}\vspace{0.0cm}

The discoveries of $^{78}$Cu and $^{79}$Cu were reported by Kratz et al. in ``Neutron-rich isotopes around the \textit{r}-process `waiting-point' nuclei $^{79}_{29}$Cu$_{50}$ and $^{80}_{30}$Zn$_{50}$'' in 1991 \cite{1991Kra01}. A $^{238}$UC-graphite target was irradiated with 600 MeV protons from the CERN synchro-cyclotron and the fragments were separated and identified with the CERN ISOLDE on-line mass separator. The observation of $^{78}$Cu was not considered new quoting a previous conference proceeding \cite{1988Lun01} and the measured half-life is only listed in a table as 342(11)~ms. ``As an example, [the figure] shows a typical multiscaling curve for A = 79, after subtraction of the 2.85~s $^{79}$Ga component. Clearly, two short-lived $\beta$dn activities are seen, $^{79}$Zn with T$_{1/2} \simeq$ 1~s and the new N = 50 `waiting-point' nucleus $^{79}$Cu with T$_{1/2}$ = (188$\pm$25)~ms.'' The half-life measured for $^{78}$Cu is included in the weighted average for the currently accepted value of 335(11)~ms and the half-life obtained for $^{79}$Cu corresponds to the presently accepted value.

\subsection{$^{80}$Cu}\vspace{0.0cm}

In 1995 Engelmann et al. reported the discovery of $^{80}$Cu in ``Production and Identification of Heavy Ni Isotopes: Evidence for the Doubly Magic Nucleus $^{78}_{28}$Ni'' \cite{1995Eng01}. $^{238}$U ions were accelerated in the UNILAC and the heavy-ion synchrotron SIS at GSI to an energy of 750 A-MeV. $^{80}$Cu was produced by projectile fission, separated in-flight by the FRS and identified event-by-event by measuring magnetic rigidity, energy loss and time of flight. ``For a total dose of 10$^{13}$ U ions delivered in 132 h on the target three events can be assigned to the isotope $^{78}$Ni. Other new nuclei, $^{77}$Ni, $^{73,74,75}$Co and $^{80}$Cu can be identified, the low count rate requires a background-free measurement.'' A total of four events of $^{80}$Cu were observed.

\section{Summary}
The discoveries of the known chromium, manganese, nickel, and copper isotopes have been compiled and the methods of their production discussed.

With a few exceptions the discovery of the chromium isotopes was straightforward. The half-life measurement of $^{46}$Cr was initially incorrect and the half-life of $^{47}$Cr did at first not have a firm element nor mass assignment. It took about 15 years following the first reports to determine and assign the correct half-life of $^{55}$Cr. Searches for $^{56}$Cr were initially unsuccessful.

The limit for observing long lived manganese isotopes beyond the proton dripline which can be measured by implantation decay studies has been reached with the discovery of $^{46}$Mn and the determination of upper limits for the half-lives of 105~ns and 70~ns for $^{44}$Mn and $^{45}$Mn, respectively. The discovery of the manganese isotopes was fairly straightforward. The half-lives of $^{49}$Mn, $^{50}$Mn, and $^{51}$Mn had been measured initially without a firm mass assignment.

The identification for most nickel isotopes was straightforward. The few exceptions included the tentative claim that $^{56}$Ni was stable, and an incorrect half-life measurement for $^{67}$Ni which was corrected only 18 years later. Also, the half-life of $^{65}$Ni was initially assigned to $^{63}$Ni.

The predicted proton dripline has been reached for copper and it might be possible to observe two additional long-lived isotopes beyond the proton dripline. The identification of the proton rich isotopes has been difficult. The half-lives of $^{61}$Cu and $^{66}$Cu had been measured initially without a firm mass assignment. The first half-life measurements of $^{58}$Cu and $^{60}$Cu were incorrect and the half-life of $^{59}$Cu was at first assigned to $^{58}$Cu. In addition, $^{62}$Cu, $^{64}$Cu and $^{66}$Cu were initially incorrectly identified as stable isotopes.

\ack

The main research on the individual elements was performed by KG (manganese and copper) and RR (chromium and nickel). This work was supported by the National Science Foundation under grants No. PHY06-06007 (NSCL).

\bibliography{../isotope-discovery-references}

\newpage

\newpage

\TableExplanation

\bigskip
\renewcommand{\arraystretch}{1.0}

\section*{Table 1.\label{tbl1te} Discovery of chromium, manganese, nickel, and copper isotopes }
\begin{tabular*}{0.95\textwidth}{@{}@{\extracolsep{\fill}}lp{5.5in}@{}}
\multicolumn{2}{p{0.95\textwidth}}{ }\\

Isotope & Chromium, Manganese, Nickel, or Copper Isotope \\
Author & First author of refereed publication \\
Journal & Journal of publication \\
Ref. & Reference \\
Method & Production method used in the discovery: \\
 & FE: fusion evaporation \\
 & LP: light-particle reactions (including neutrons) \\
 & MS: mass spectroscopy \\
 & PN: Photo Nuclear Reactions \\
 & NC: neutron capture reactions \\
 & NF: Neutron Induced Fission \\
 & CPF: Charged-Particle Induced Fission \\
 & TR: Heavy-Ion Transfer \\
 & DI: deep-inelastic reactions \\
 & PF: projectile fragmentation or fission \\
 & SP: spallation reactions \\
Laboratory & Laboratory where the experiment was performed\\
Country & Country of laboratory\\
Year & Year of discovery \\
\end{tabular*}
\label{tableI}

\datatables 



\setlength{\LTleft}{0pt}
\setlength{\LTright}{0pt}


\setlength{\tabcolsep}{0.5\tabcolsep}

\renewcommand{\arraystretch}{1.0}

\footnotesize 

\begin{longtable}{@{\extracolsep\fill}llllllll@{}}
\caption{Discovery of Chromium, Manganese, Nickel, and Copper Isotopes. See page\ \pageref{tbl1te} for Explanation of Tables}
Isotope & First Author & Journal & Ref. & Method & Laboratory & Country & Year\\
\hline\\
\endfirsthead\\
\caption[]{(continued)}
Isotope & First Author & Journal & Ref. & Method & Laboratory & Country & Year\\
\hline\\
\endhead
$^{42}$Cr   & B. Blank & Phys. Rev. Lett. &\cite{1996Bla01}& PF & Darmstadt & Germany &1996 \\
$^{43}$Cr   & V. Borrel & Z. Phys. A &\cite{1992Bor01}& PF & GANIL & France &1992 \\
$^{44}$Cr   & F. Pougheon & Z. Phys. A &\cite{1987Pou01}& PF & GANIL & France &1987 \\
$^{45}$Cr   & K.P. Jackson & Phys. Lett. B &\cite{1974Jac01}& FE & Chalk River & Canada &1974 \\
$^{46}$Cr   & J. Zioni & Nucl. Phys. A &\cite{1972Zio01}& FE & Jerusalem & Israel &1972 \\
$^{47}$Cr   & I.D. Proctor & Phys. Rev. Lett. &\cite{1972Pro01}& LP & Michigan State & USA &1972 \\
$^{48}$Cr   & G. Rudstam & Phys. Rev. &\cite{1952Rud01}& SP & Berkeley & USA &1952 \\
$^{49}$Cr   & J.J. O'Connor & Phys. Rev. &\cite{1942OCo01}& LP & Ohio State & USA &1942 \\
$^{50}$Cr   & F.W. Aston & Nature &\cite{1930Ast03}& MS & Cambridge & UK &1930 \\
$^{51}$Cr   & H. Walke & Phys. Rev. &\cite{1940Wal03}& LP & Berkeley & USA &1940 \\
$^{52}$Cr   & F.W. Aston & Nature &\cite{1923Ast01}& MS & Cambridge & UK &1923 \\
$^{53}$Cr   & F.W. Aston & Nature &\cite{1930Ast03}& MS & Cambridge & UK &1930 \\
$^{54}$Cr   & F.W. Aston & Nature &\cite{1930Ast03}& MS & Cambridge & UK &1930 \\
$^{55}$Cr   & A. Flammersfeld & Z. Naturforsch. &\cite{1952Fla01}& NC & Mainz & Germany &1952 \\
$^{56}$Cr   & B.J. Dropesky & Nucl. Phys. &\cite{1960Dro01}& LP & Los Alamos & USA &1960 \\
$^{57}$Cr   & C.N. Davids & Phys. Rev. C &\cite{1978Dav01}& FE & Argonne & USA &1978 \\
$^{58}$Cr   & D. Guerreau & Z. Phys. A &\cite{1980Gue01}& DI & Orsay & France &1980 \\
$^{59}$Cr   & D. Guerreau & Z. Phys. A &\cite{1980Gue01}& DI & Orsay & France &1980 \\
$^{60}$Cr   & H. Breuer & Phys. Rev. C &\cite{1980Bre01}& DI & Berkeley & USA &1980 \\
            & U. Bosch & Nucl. A &\cite{1988Bos01}& DI & Darmstadt & Germany &1988 \\
$^{61}$Cr   & D. Guillemaud-Mueller & Z. Phys. A &\cite{1985Gui01}& PF & GANIL & France &1985 \\
$^{62}$Cr   & D. Guillemaud-Mueller & Z. Phys. A &\cite{1985Gui01}& PF & GANIL & France &1985 \\
$^{63}$Cr   & M. Weber & Z. Phys. A &\cite{1992Web01}& PF & Darmstadt & Germany &1992 \\
$^{64}$Cr & M. Weber & Z. Phys. A &\cite{1992Web01}& PF & Darmstadt & Germany &1992 \\
$^{65}$Cr & M. Bernas & Phys. Lett. B &\cite{1997Ber01}& PF & Darmstadt & Germany &1997 \\
$^{66}$Cr & M. Bernas & Phys. Lett. B &\cite{1997Ber01}& PF & Darmstadt & Germany &1997 \\
$^{67}$Cr & M. Bernas & Phys. Lett. B &\cite{1997Ber01}& PF & Darmstadt & Germany &1997 \\
$^{68}$Cr & O.B. Tarasov & Phys. Rev. Lett. &\cite{2009Tar01}& PF & Michigan State & USA &2009 \\
 &  &  &  &  &  & & \\
 &  &  &  &  &  & & \\
$^{46}$Mn   & F. Pougheon & Z. Phys. A &\cite{1987Pou01}& PF & GANIL & France &1987 \\
$^{47}$Mn   & F. Pougheon & Z. Phys. A &\cite{1987Pou01}& PF & GANIL & France &1987 \\
$^{48}$Mn   & T. Sekine & Nucl. Phys. A &\cite{1987Sek01}& FE & Darmstadt & Germany &1987 \\
$^{49}$Mn   & J. Cerny & Phys. Rev. Lett. &\cite{1970Cer02}& FE & Oxford & UK &1970 \\
$^{50}$Mn   & W.M. Martin & Can. J. Phys. &\cite{1952Mar01}& LP & McGill & Canada &1952 \\
$^{51}$Mn   & J.J. Livingood & Phys. Rev. &\cite{1938Liv02}& LP & Berkeley & USA &1938 \\
$^{52}$Mn   & J.J. Livingood & Phys. Rev. &\cite{1938Liv02}& LP & Berkeley & USA &1938 \\
$^{53}$Mn   & J.R. Wilkinson & Phys. Rev. &\cite{1955Wil01}& LP & Berkeley & USA &1955 \\
$^{54}$Mn   & J.J. Livingood & Phys. Rev. &\cite{1938Liv02}& LP & Berkeley & USA &1938 \\
$^{55}$Mn   & F.W. Aston & Nature &\cite{1923Ast01}& MS & Cambridge & UK &1923 \\
$^{56}$Mn   & E. Amaldi & Ric. Scientifica &\cite{1934Ama01}& NC & Rome & Italy &1934 \\
$^{57}$Mn   & B.L. Cohen & Phys. Rev. &\cite{1954Coh01}& LP & Oak Ridge & USA &1954 \\
$^{58}$Mn   & D.M. Chittenden & Phys. Rev. &\cite{1961Chi01}& LP & Arkansas & USA &1961 \\
$^{59}$Mn   & E. Kashy & Phys. Rev. C &\cite{1976Kas01}& LP & Michigan State & USA &1976 \\
$^{60}$Mn   & E.B. Norman & Phys. Rev. C &\cite{1978Nor01}& FE & Argonne & USA &1978 \\
$^{61}$Mn   & D. Guerreau & Z. Phys. A &\cite{1980Gue01}& DI & Orsay & France &1980 \\
$^{62}$Mn   & E. Runte & Nucl. Phys. A & \cite{1983Run01} & DI & Darmstadt & Germany &1983 \\
$^{63}$Mn   & U. Bosch & Phys. Lett. B &\cite{1985Bos01}& DI & Darmstadt & Germany &1985 \\
$^{64}$Mn & D. Guillemaud-Mueller & Z. Phys. A &\cite{1985Gui01}& PF & GANIL & France &1985 \\
$^{65}$Mn & D. Guillemaud-Mueller & Z. Phys. A &\cite{1985Gui01}& PF & GANIL & France &1985 \\
$^{66}$Mn & M. Weber & Z. Phys. A &\cite{1992Web01}& PF & Darmstadt & Germany &1992 \\
$^{67}$Mn & M. Bernas & Phys. Lett. B &\cite{1997Ber01}& PF & Darmstadt & Germany &1997 \\
 &  &  &  &  &  & & \\
 &  &  &  &  &  & & \\
$^{48}$Ni & B. Blank & Phys. Rev. Lett. & \cite{2000Bla01} & PF & GANIL & France &2000 \\
$^{49}$Ni & B. Blank & Phys. Rev. Lett. & \cite{1996Bla01} & PF & Darmstadt & Germany &1996 \\
$^{50}$Ni & B. Blank & Phys. Rev. C & \cite{1994Bla01} & PF & Darmstadt & Germany &1994 \\
$^{51}$Ni & F. Pougheon & Z. Phys. A & \cite{1987Pou01} & PF & GANIL & France &1987 \\
$^{52}$Ni & F. Pougheon & Z. Phys. A & \cite{1987Pou01} & PF & GANIL & France &1987 \\
$^{53}$Ni & D.J. Vieira & Phys. Lett. B & \cite{1976Vie01} & FE & Berkeley & USA &1976 \\
$^{54}$Ni & R.E. Tribble & Phys. Rev. C & \cite{1977Tri02} & LP & Texas A\&M& USA &1977 \\
$^{55}$Ni & I.D. Proctor & Phys. Rev. Lett. & \cite{1972Pro01} & LP & Michigan State & USA &1972 \\
$^{56}$Ni & W.J. Worthington & Phys. Rev. & \cite{1952Wor01} & LP & Berkeley & USA &1952 \\
$^{57}$Ni & J.J. Livingood & Phys. Rev. & \cite{1938Liv04} & LP & Berkeley & USA &1938 \\
$^{58}$Ni & F.W. Aston & Nature & \cite{1921Ast02} & MS & Cambridge & UK &1921 \\
$^{59}$Ni & A.R. Brosi & Phys. Rev. & \cite{1951Bro01} & NC & Oak Ridge & USA &1951 \\
$^{60}$Ni & F.W. Aston & Nature & \cite{1921Ast02} & MS & Cambridge & UK &1921 \\
$^{61}$Ni & F.W. Aston & Nature & \cite{1934Ast01} & MS & Cambridge & UK &1934 \\
$^{62}$Ni & F.W. Aston & Nature & \cite{1934Ast01} & MS & Cambridge & UK &1934 \\
$^{63}$Ni & A.R. Brosi & Phys. Rev. & \cite{1951Bro01} & NC & Oak Ridge & USA &1951 \\
$^{64}$Ni & J. de Gier &  Proc. Akad. Soc. & \cite{1935deG02} & MS & Amsterdam & Netherlands &1935 \\
$^{65}$Ni & J.A. Swartout & Phys. Rev. & \cite{1946Swa01} & LP & Oak Ridge & USA &1946 \\
$^{66}$Ni & R.H. Goeckermann & Phys. Rev. & \cite{1948Goe01} & CPF & Berkeley & USA &1948 \\
$^{67}$Ni & R. T.Kouzes & Phys. Rev. C & \cite{1978Kou02} & LP & Princeton & USA &1978 \\
$^{68}$Ni & T.S. Bhatia & Z. Phys. A & \cite{1977Bha01} & TR & Heidelberg & Germany &1977 \\
$^{69}$Ni & Ph. Dessagne & Nucl. Phys. A & \cite{1984Des01} & TR & Orsay & France &1984 \\
$^{70}$Ni & P. Armbruster & Europhys. Lett. & \cite{1987Arm01} & NF & Grenoble & France &1987 \\
$^{71}$Ni & P. Armbruster & Europhys. Lett. & \cite{1987Arm01} & NF & Grenoble & France &1987 \\
$^{72}$Ni & P. Armbruster & Europhys. Lett. & \cite{1987Arm01} & NF & Grenoble & France &1987 \\
$^{73}$Ni & P. Armbruster & Europhys. Lett. & \cite{1987Arm01} & NF & Grenoble & France &1987 \\
$^{74}$Ni & P. Armbruster & Europhys. Lett. & \cite{1987Arm01} & NF & Grenoble & France &1987 \\
$^{75}$Ni & M. Weber & Z. Phys. A & \cite{1992Web01} & PF & Darmstadt & Germany &1992 \\
$^{76}$Ni & Ch. Engelmann & Z. Phys. A & \cite{1995Eng01}& PF & Darmstadt & Germany &1995 \\
$^{77}$Ni & Ch. Engelmann & Z. Phys. A & \cite{1995Eng01}& PF & Darmstadt & Germany &1995 \\
$^{78}$Ni & Ch. Engelmann & Z. Phys. A & \cite{1995Eng01}& PF & Darmstadt & Germany &1995 \\
 &  &  &  &  &  & & \\
 &  &  &  &  &  & & \\
$^{55}$Cu & F. Pougheon & Z. Phys. A & \cite{1987Pou01} & PF & GANIL & France &1987 \\
$^{56}$Cu & F. Pougheon & Z. Phys. A & \cite{1987Pou01} & PF & GANIL & France &1987 \\
$^{57}$Cu & D.J. Vieira & Phys. Lett. B & \cite{1976Vie01} & FE & Berkeley & USA &1976 \\
$^{58}$Cu & W.M. Martin & Can. J. Phys. & \cite{1952Mar01} & LP & McGill & Canada &1952 \\
$^{59}$Cu & C.E. Leith & Phys. Rev. & \cite{1947Lei01} & LP & Berkeley & USA &1947 \\
          & L. Lindner & Physica & \cite{1955Lin02} & LP &  & Amsterdam &1955 \\
$^{60}$Cu & C.E. Leith & Phys. Rev. & \cite{1947Lei01} & LP & Berkeley & USA &1947 \\
$^{61}$Cu & L.N. Ridenour & Phys. Rev. & \cite{1937Rid01} & LP & Princeton & USA &1937 \\
$^{62}$Cu & F.A. Heyn & Nature & \cite{1936Hey01} & LP & Eindhoven & Netherlands &1936 \\
$^{63}$Cu & F.W. Aston & Nature & \cite{1923Ast02} & MS & Cambridge & UK &1923 \\
$^{64}$Cu & S.N. Van Voorhis & Phys. Rev. & \cite{1936Voo01} & LP & Berkeley & USA &1936 \\
$^{65}$Cu & F.W. Aston & Nature & \cite{1923Ast02} & MS & Cambridge & UK &1923 \\
$^{66}$Cu & W.Y. Chang & Nature & \cite{1937Cha01} & LP & Cambridge & UK &1937 \\
$^{67}$Cu & R.H. Goeckermann & Phys. Rev. & \cite{1948Goe01} & CPF & Berkeley & USA &1948 \\
$^{68}$Cu & A. Flammersfeld & Z. Naturforsch. & \cite{1953Fla01} & LP & Mainz & Germany &1953 \\
$^{69}$Cu & J. van Klinken & Phys. Rev. & \cite{1966Kli01} & PN & Iowa State & USA &1966 \\
$^{70}$Cu & L.M. Taff & Nucl. Phys. A & \cite{1971Taf01} & LP & Oak Ridge & USA &1971 \\
$^{71}$Cu & E. Runte & Nucl. Phys. A & \cite{1983Run01} & DI & Darmstadt & Germany &1983 \\
$^{72}$Cu & E. Runte & Nucl. Phys. A & \cite{1983Run01} & DI & Darmstadt & Germany &1983 \\
$^{73}$Cu & E. Runte & Nucl. Phys. A & \cite{1983Run01} & DI & Darmstadt & Germany &1983 \\
$^{74}$Cu & P. Armbruster & Europhys. Lett. & \cite{1987Arm01} & NF & Grenoble & France &1987 \\
$^{75}$Cu & P.L. Reeder& Phys. Rev. C & \cite{1985Ree01} & NF & Brookhaven & USA &1985 \\
$^{76}$Cu & P. Armbruster & Europhys. Lett. & \cite{1987Arm01} & NF & Grenoble & France &1987 \\
$^{77}$Cu & P. Armbruster & Europhys. Lett. & \cite{1987Arm01} & NF & Grenoble & France &1987 \\
$^{78}$Cu & K.-L. Kratz & Z. Phys. A & \cite{1991Kra01} & SP & CERN & Switzerland &1991 \\
$^{79}$Cu & K.-L. Kratz & Z. Phys. A & \cite{1991Kra01} & SP & CERN & Switzerland &1991 \\
$^{80}$Cu & Ch. Engelmann & Z. Phys. A & \cite{1995Eng01}& PF & Darmstadt & Germany &1995 \\

\end{longtable}

\end{document}